\documentclass[final,3p,times,authoryear]{elsarticle}
\usepackage{amssymb}
\usepackage{amsmath}
\usepackage{lineno}
\usepackage{graphicx}
\usepackage{caption}
\usepackage{makecell}
\usepackage{hyperref}
\hypersetup{colorlinks=true,urlcolor=blue}

\journal{Communications in Nonlinear Science and Numerical Simulation}

\begin{document}

\begin{frontmatter}

\title{A Mathematical Multiscale Framework for Nonlinear Modeling of the Heliosphere}

\author[aff1]{Daniele Telloni\corref{cor1}\fnref{same}}
\author[aff2,aff1]{Sabrina Guastavino\fnref{same}}
\author[aff2,aff1]{Michele Piana}

\affiliation[aff1]{organization={National Institute for Astrophysics, Astrophysical Observatory of Torino},
                        addressline={Via Osservatorio 20},
                        city={Pino Torinese},
                        postcode={10025},
                        country={Italy}}
\affiliation[aff2]{organization={Department of Mathematics, University of Genoa},
                        addressline={Via Dodecaneso 35},
                        city={Genoa},
                        postcode={16146},
                        country={Italy}}

\cortext[cor1]{Corresponding author: Daniele Telloni (\href{daniele.telloni@inaf.it}{daniele.telloni@inaf.it})}
\fntext[same]{These authors contributed equally to this work.}

\begin{abstract}
	The inner heliosphere is a highly nonlinear continuous medium where macroscopic ideal magnetohydrodynamics is inextricably coupled with microscopic, fully developed stochastic turbulence. State-of-the-art global $3$D numerical solvers, while proficient at capturing the large-scale propagation of extreme space physics phenomena such as coronal mass ejections and corotating interaction regions, are computationally constrained by grid resolution and fundamental partial differential equation formulations. Consequently, they inherently do not resolve the local small-scale turbulent cascade, magnetic intermittency, and localized thermodynamic anomalies. To bridge this fundamental scale-separation gap, this paper presents a novel mathematical and numerical framework that harmoniously integrates global heliospheric structures with local turbulent physics. The framework generates a highly realistic synthetic continuous medium by superimposing a stochastically driven nonlinear turbulent cascade onto a bimodal deterministic background flow. This cascade reproduces theoretical constraints for fractional power spectral densities, magnetic compressibility, and small-scale intermittency via a discrete multiplicative binomial cascade (p-model). This multiscale continuous background is dynamically perturbed by analytical models of large-scale nonlinear transients, including kinematic shock steepening, continuous current sheet topological reversals, and solar ejecta formalized as asymmetric force-free magnetic flux ropes. By generating exact spatio-temporal arrays across arbitrary distributed coordinates, this unified mathematical model overcomes the spatial degeneracy of single-point observations and provides a computationally efficient analytical baseline for performance models supporting future in-situ multi-spacecraft heliospheric missions, which are currently in their feasibility study phase.
\end{abstract}

\begin{highlights}
	\item A novel multiscale mathematical framework is introduced to seamlessly couple analytical solutions of large-scale heliospheric structures with a stochastic multifractal turbulent cascade.
	\item The stochastic turbulent manifold enforces fractional power-law spectra and non-Gaussian intermittency through the spectral filtering of a multifractal singular measure.
	\item Large-scale nonlinear transients are formally modeled utilizing force-free Lundquist fields, Rankine-Hugoniot kinematic steepening, and continuous $\mathcal{C}^{\infty}$ operators for topological boundary reversals.
	\item By generating synthetic time series across arbitrary distributed spatial grids, the framework breaks single-point spatial degeneracy to validate future multi-spacecraft mission concepts.
\end{highlights}

\begin{keyword}
	Multiscale deterministic-stochastic modeling \sep Nonlinear ideal magnetohydrodynamics \sep Multifractal singular measures \sep Fractional spectral synthesis \sep Continuous topological operators \sep Eulerian multi-point projections
	\PACS 05.10.-a \sep 95.75.Wx \sep 05.45.Tp \sep 02.60.-x \sep 95.75.Pq \sep 52.35.Ra \sep 94.05.Lk \sep 47.27.E \sep 52.30.Cv \sep 96.50.Ci \sep 96.60.ph
	\MSC[2020] 65C20 \sep 65T50 \sep 76F55 \sep 76M35 \sep 76W05
\end{keyword}

\end{frontmatter}

\section{Introduction}
\label{sec:introduction}
The inner heliosphere behaves as a highly complex, collisionless, and magnetized fluid exhibiting fully developed nonlinear turbulence. Large-scale nonlinear structures within this continuous medium, particularly propagating Coronal Mass Ejections (CMEs) and Corotating Interaction Regions (CIRs), drive severe geomagnetic storms that threaten modern technological infrastructure \citep{2022FrASS...917103B}. The mathematical modeling of this environment presents a formidable multiscale challenge: the physical domain spans from the macroscopic evolution of massive $3$D topological structures (spanning tens of millions of kilometers) to the microscopic dissipation scales of the turbulent cascade (tens of kilometers) \citep{1995SSRv...73....1T}. As a consequence, the operational attempt to predict the exact $3$D topology and Time of Arrival (ToA) of these structures based on single-point temporal observations (such as those provided by L$1$ monitors) represents a severely ill-posed inverse problem \citep{2015SunGe..10..119M}. Extracting the $3$D curvature, propagation vector, and true magnetic chirality of a massive transient from a $1$D Eulerian temporal cut is mathematically degenerate. Breaking this spatial degeneracy requires the transition to multi-point spatial sampling \citep{2000JASTP..62.1251S,2015JSWSC...5A...3R}.

Currently, the computational paradigm for heliospheric forecasting is based on sophisticated global $3$D MagnetoHydroDynamic (MHD) codes, such as Enlil \citep{2003AdSpR..32..497O}, the EUropean Heliospheric FORecasting Information Asset \citep[EUHFORIA;][]{2018JSWSC...8A..35P}, and Reverse In situ data and MHD APproach \citep[EUHFORIA;][]{2021JSWSC..11....7B}. These deterministic models utilize computationally intensive finite-difference or finite-volume schemes to solve nonlinear partial differential equations on coarse numerical grids. Although exceptionally capable of describing the large-scale Parker spiral structure and the macroscopic trajectory of transient events, these Eulerian solvers are fundamentally penalized by numerical dissipation and grid resolution limits. They inherently filter out the sub-grid microscopic physics, failing to resolve the stochastic turbulent cascade, the localized non-Gaussian intermittent structures, and the micro-scale thermodynamic anomalies that are ubiquitous in the in-situ observations of the solar wind. From a mathematical physics perspective, this lack of local stochasticity fundamentally limits their capability to accurately predict critical local parameters, such as the Alfv\'enicity of wind streams \citep[known to play an important role in the recovery phase of geomagnetic storms, e.g.,][]{2021ApJ...916...64T} or the exact small-scale variance of the southward magnetic field component ($B_{z}$), which ultimately dictates the geo-effectiveness of the transient through magnetic reconnection \citep{1961PhRvL...6...47D}.

The main contribution of this work is the development of a unified deterministic-stochastic numerical framework for modeling large-scale heliospheric structures within an intermittent turbulent solar wind. The proposed methodology integrates three complementary components: (i) an analytical representation of the large-scale solar wind based on Parker-spiral geometry and radial plasma expansion; (ii) a stochastic multifractal turbulence model with controllable spectral, intermittent, and covariance properties, dynamically conditioned on the local solar wind regime; and (iii) a set of nonlinear $\mathcal{C}^{\infty}$ topological operators describing CMEs, CIRs, and Heliospheric Current Sheet (HCS) crossings, together with their associated modifications of the local turbulent statistics. By combining these components within a common numerical formulation, the framework enables the generation of spatially coherent macroscopic structures embedded in statistically controlled turbulent backgrounds, providing synthetic multipoint observations for numerical experimentation and the evaluation of heliospheric reconstruction methods.

The paper is structured as follows. \S{} \ref{sec:synthetic_environment} details the mathematical formulation of the synthetic background space environment, coupling the stationary ideal MHD equations with the stochastic injection of a turbulent cascade via a localized multiplicative binomial cascade \citep[p-model,][]{1987PhRvL..59.1424M}. \S{} \ref{sec:transients} extends the framework by formalizing the analytical, nonlinear mapping of large-scale transients. \S{} \ref{sec:validation} provides the multi-point validation of the generated fields, demonstrating both the morphological realism of the simulated macroscopic transients and the analytical control over fractional spectral envelopes and non-Gaussian small-scale intermittency. Finally, \S{} \ref{sec:conclusions} outlines the broader implications of the present mathematical modeling approach for the future of computational space physics and the interpretation of distributed multi-point observations.

\section{Mathematical formulation of the synthetic interplanetary space environment}
\label{sec:synthetic_environment}
To evaluate the performance of distributed multi-point configurations and their associated operational algorithms, the proposed simulation framework must first construct a mathematically coherent representation of the interplanetary medium. This section details the continuous formulations employed to generate the macroscopic solar wind background and the stochastic injection of a physically consistent turbulent cascade.

\subsection{Stationary MHD solutions and macroscopic radial scalings}
\label{subsec:solar_wind_background}
The simulation is initialized by defining the large-scale properties of the background solar wind at a reference boundary distance of $r_{0}=1$ Astronomical Unit (au). The state of the solar wind is modeled as a bimodal velocity distribution, defining either slow, dense, and highly compressible streams, or fast, tenuous, and highly Alfv\'enic streams \citep{2013LRSP...10....2B}.

The background magnetic field $\mathbf{B}$ is formally structured according to the classic stationary Parker spiral solution \citep{1958ApJ...128..664P}. For an observer located at a heliocentric radial distance $r$ and heliolongitude $\phi$, the spiral angle $\psi$ represents the interplay between the solar rotation rate $\Omega_{\odot}$ and the local macroscopic radial bulk speed $V$:

\begin{equation}
	\psi=\arctan\left(\frac{\Omega_{\odot}r}{V}\right).
\end{equation}

The macroscopic fields at $1$ au are subsequently back-projected to the specific observational coordinates ($r_{node},\phi_{node}$) of each arbitrary spatial node within the observational domain. This spatial mapping is constrained by the exact conservation laws of stationary ideal MHD. Assuming a spherically symmetric, time-independent expansion, the solenoidal condition for the magnetic field ($\nabla\cdot\mathbf{B}=0$) dictates the spatial decay of the radial component $B_{r}$:

\begin{equation}
	\frac{1}{r^{2}}\frac{\partial}{\partial r}\left(r^{2}B_{r}\right)=0\quad\implies\quad B_{r}(r)=B_{r,0}\left(\frac{r_{0}}{r}\right)^{2},
\end{equation}

\noindent where the subscript $0$ denotes the reference value evaluated at $r_{0}$. The plasma mass density $\rho$ follows the exact same $r^{-2}$ scaling, dictated by the stationary continuity equation ($\nabla\cdot(\rho\mathbf{V})=0$) under the assumption of a constant radial velocity vector $\mathbf{V}$.

Conversely, the azimuthal component of the Interplanetary Magnetic Field (IMF), $B_{\phi}$, is governed by the steady induction equation $\nabla\times(\mathbf{V}\times\mathbf{B})=\mathbf{0}$, reflecting the frozen-in flux theorem. The ratio of the azimuthal to radial magnetic field components in the solar equatorial plane is proportional to the ratio of the corotation velocity to the radial expansion speed. Utilizing the solar angular velocity $\Omega_{\odot}$, this stationary relationship yields:

\begin{equation}
	\frac{B_{\phi}}{B_{r}}=\frac{-\Omega_{\odot}r}{V}\quad\implies\quad B_{\phi}(r)=-B_{r,0}\left(\frac{r_{0}}{r}\right)^{2}\frac{\Omega_{\odot}r}{V},
\end{equation}

\noindent which explicitly demonstrates that $B_{\phi}$ scales inversely with distance as $r^{-1}$.

To close the fluid equations thermodynamically, the macroscopic proton temperature $T$ must be specified. In the solar wind, $T$ is governed by a polytropic equation of state $P\propto\rho^{\gamma}\implies T\propto\rho^{\gamma-1}$, yielding a spatial profile:

\begin{equation}
	T(r)=T_{0}\left(\frac{r_{0}}{r}\right)^{2(\gamma-1)}.
\end{equation}

Decades of statistical parameter estimation from inner heliosphere observations \citep{1995JGR...100...13T,1995GeoRL..22..325R,2011JGRA..116.9105H} empirically indicate a non-adiabatic temperature $r^{-0.8}$ scaling, which corresponds to an effective polytropic index $\gamma\approx1.4$.

To continuously model the temporal evolution of the time series and the formation of Stream Interaction Regions (SIRs) or CIRs across the computational domain, the transition between the asymptotic slow and fast solar wind regimes is mathematically formalized via a dimensionless transition operator $\chi:\mathbb{R}\to[0,1]$. To guarantee $\mathcal{C}^\infty$ continuity across the stream interface, $\chi(t)$ is defined as a hyperbolic tangent function:

\begin{equation}
	\chi(t)=\frac{1}{2}\left[1+\tanh\left(\frac{t-t_{s}}{w_{s}}\right)\right],
\end{equation}

\noindent where $t_{s}$ denotes the randomly initialized arrival time of the stream interface and $w_{s}$ parametrizes the temporal width of the velocity shear layer. Any macroscopic stationary variable $U(t)$, encompassing the bulk speed $V$, plasma mass density $\rho$, temperature $T$, Alfv\'enicity $r_{A}$, and magnetic compressibility $C_{B}$, is dynamically evaluated at each discrete time step as a convex combination of its nominal slow ($U_{S}$) and fast ($U_{F}$) asymptotic states:

\begin{equation}
	U(t)=U_{S}[1-\chi(t)]+U_{F}\chi(t).
\end{equation}

Consequently, to strictly preserve the additivity of variances for independent stochastic processes, the zero-mean fluctuating turbulent fields (such as the generic magnetic perturbation $\delta\mathbf{b}$) are synthesized by weighting the regime-specific fractional cascades with the square root of the transition operator:

\begin{equation}
	\delta\mathbf{b}(t)=\delta\mathbf{b}_{F}(t)\sqrt{\chi(t)}+\delta\mathbf{b}_{S}(t)\sqrt{1-\chi(t)},
\end{equation}

where $\delta\mathbf{b}_F$ and $\delta\mathbf{b}_S$ denote the magnetic perturbations associated to fast and slow states, respectively.

Furthermore, the dynamic pile-up of plasma and magnetic flux ahead of the high-speed stream is mathematically modeled by introducing a nonlinear compression operator $\Gamma:\mathbb{R}\to\mathbb{R}$. Given a theoretical shock compression ratio $r_{c}$, the macroscopic amplification across the SIR/CIR is governed by:

\begin{equation}
	\Gamma(t)=1\cdot[1-\chi(t)]+r_{c}\chi(t).
\end{equation}

This strictly positive operator $\Gamma(t)$ is applied multiplicatively to the transverse components of the macroscopic induction field ($B_{y},B_{z}$) and to the thermodynamic scalar fields ($\rho,T$). This algebraic formulation effectively models the nonlinear steepening of the compression region into a forward shock structure, ensuring mass and transverse flux conservation as the interaction region propagates radially outward.

Finally, temporal phase shifting is applied to account for the radial advection of the plasma parcels from the spatial node to the $1$ au reference boundary, incorporating co-rotating longitudinal delays:

\begin{equation}
	t_{node}=t_{1au}-\frac{(1-r_{node})L_{au}}{V}+\frac{\phi_{node}}{\Omega_{\odot}},
\end{equation}

where $t_{node}$ and $t_{1au}$ represent the continuous temporal coordinates at the node's heliocentric distance ($r_{node}$, in astronomical units) and at $1$ au, respectively, and $L_{au}\approx1.5\times10^{8}$ km is the astronomical unit conversion constant.

It is crucial to note that because the macroscopic velocity field $V(t)$ is explicitly time-dependent, the advective mapping $t_{1au}\mapsto t_{node}$ is inherently nonlinear, resulting in a distorted, non-uniform temporal grid at the observer's location. To ensure numerical consistency for the subsequent spectral processing, the advected time array is monotonically sorted, and the macroscopic scalar and vector fields are numerically remapped onto the uniform discrete computational grid $t_{k}=k\Delta t$, with $k\in\mathbb{N}$ and temporal sampling interval $\Delta t$, via a linear interpolation scheme.

\subsection{Stochastic generation of the turbulent cascade and spectral shaping}
\label{subsec:turbulence_injection}
The interplanetary medium is a natural laboratory for fully developed, collisionless MHD turbulence \citep{2013LRSP...10....2B}. Synthetic continuous fields consisting solely of macroscopic stationary solutions fail to capture the multiscale nonlinear dynamics required to properly constrain high-frequency local fluctuations and multiscale statistical properties. Therefore, a synthetic stochastic cascade is dynamically superimposed onto the macroscopic deterministic background.

The mathematical generation of realistic turbulent fluctuations relies on the convolution of a multifractal singular measure with a fractional integration scheme in the Fourier domain. To inject the required spatial intermittency, a hallmark of nonlinear cascading processes where the active volume of turbulence strictly decreases at smaller scales, a discrete binomial multiplicative cascade, known as the p-model, is implemented \citep{1987PhRvL..59.1424M}. Let $L$ be the integral scale of the domain and $N$ the total number of discrete grid points. The cascade is constructed over $n=\log_{2}(N)$ iterative steps. At the $j$-th step ($j=1,\dots,n$), the domain is partitioned into $2^{j}$ segments of length $l_{j}=L/2^{j}$. The localized spatial energy dissipation rate, defined as a probability measure $\mu$, is redistributed iteratively. If $\mu_{k}^{(j)}$ represents the measure of the $k$-th segment at step $j$, the measure is fragmented into two adjacent sub-segments at step $j+1$ via stochastic multipliers $m_{1}$ and $m_{2}$:

\begin{equation}
	\mu_{2k-1}^{(j+1)}=m_{1}\mu_{k}^{(j)},\quad\mu_{2k}^{(j+1)}=m_{2}\mu_{k}^{(j)},
\end{equation}

\noindent where the weights are uniquely determined by the intermittency parameter $p\in[0.5,1.0)$

\begin{equation}
	m_{1}=2p,\quad m_{2}=2(1-p).
\end{equation}

The assignment of $m_{1}$ and $m_{2}$ to the left or right sub-segment is treated as an independent Bernoulli trial with probability $P=1/2$. The parameter $p$ governs the generalized fractal dimensions $D_{q}$ of the resulting singular measure. Setting $p=0.5$ yields a space-filling, homogeneous (Gaussian) field with $D_{q}=1$ for all $q$. Conversely, $p>0.5$ generates highly singular, non-Gaussian structures mapping the topological distribution of intermittent current sheets and coherent vortices \citep{1999PPCF...41A.787V}.

A localized stochastic increment $\delta\xi(t)$ is subsequently generated by modulating a standard Gaussian white noise process $\eta(t)\sim\mathcal{N}(0,1)$ with the square root of the finalized spatial measure $\mu^{(n)}(t)$:

\begin{equation}
	\delta\xi(t)=\eta(t)\sqrt{\mu^{(n)}(t)}.
\end{equation}

To impose the correct two-point correlation statistics (i.e., the spectral energy distribution), the intermittent spatial field $\delta\xi(t)$ is mapped into the frequency domain via the discrete Fourier transform operator $\mathcal{F}$. The spectral amplitudes are deterministically constrained by a theoretical fractional power-law envelope, $P(f)$ in the frequency domain $f$.


Inner heliospheric turbulence typically exhibits a characteristic spectral break $f_{break}$, demarcating the boundary between the $1/f$ energy injection range and the fully developed inertial range \citep{1995SSRv...73....1T}. The exact topology of the theoretical fractional power-law envelope, $P(f)$, is mathematically conditioned upon the macroscopic state of the fluid. Fast solar wind streams, originating from coronal holes, typically preserve the signature of low-frequency Alfv\'enic injection scales. Consequently, for the fast regime, a piecewise shaping function featuring a spectral break $f_{break}$ is implemented:

\begin{equation}
	P_{F}(f)=
	\begin{cases}
		f^{-1} & f<f_{break}\\
		f_{break}^{2/3}f^{-5/3} & f\ge f_{break}
	\end{cases}
\end{equation}

Conversely, the slow solar wind, having undergone extensive dynamic processing and turbulent aging, is mathematically formalized as a fully developed turbulent medium. For this regime, the $1/f$ injection range is entirely bypassed within the simulated temporal window, and the spectral envelope is governed by a strict fractional Kolmogorov scaling across all resolved non-zero frequencies:

\begin{equation}
	P_{S}(f)=f^{-5/3}.
\end{equation}

From a physical perspective, the synthesis of the cascade relies fundamentally on Taylor's frozen-in hypothesis. While the p-model inherently generates a spatial multifractal measure $\mu$ and theoretical cascades operate in the wavenumber domain $k$, the numerical framework applies the shaping envelope $P(f)$ directly in the Eulerian temporal frequency domain $f$. This mapping implicitly assumes that the macroscopic advective bulk speed highly exceeds the local Alfv\'en and sound speeds ($V\gg v_{A},c_{s}$), mapping spatial structures into temporal signals via $f\approx kV/2\pi$. By applying the fractional power laws directly in the frequency domain, the framework effectively models the continuous advection of the frozen turbulent manifold past the stationary observer frame.

The complex spectrum is element-wise multiplied by either $\sqrt{P_{F}(f)}$ or $\sqrt{P_{S}(f)}$, explicitly coupling the turbulent cascade's multifractal properties to the specific macroscopic advective conditions of the simulated heliospheric medium, and subsequently mapped back into the time domain via the inverse Fourier operator $\mathcal{F}^{-1}$. This spectral coloring acts as a fractional differential operator, yielding the finalized stochastic fluctuation component. From a computational perspective, to ensure that the resultant stochastic time series represents a physical, strictly real-valued observable, the preservation of Hermitian symmetry within the complex Fourier domain is enforced. The spectral envelope $P(f)$ is constructed as a strictly even function over the discrete frequency vector ($P(f_{k})=P(-f_{k})$). Consequently, the algebraic multiplication of the complex noise spectrum by $\sqrt{P(f)}$ inherently preserves the conjugate symmetry of the original real-valued noise. To correct for machine-precision floating-point arithmetic errors introduced during the discrete $\mathcal{F}^{-1}$ operation, a formal projection operator onto the real manifold ($\Re\{\cdot\}$) is explicitly applied

\begin{equation}
	\delta\tilde{b}(t)=\Re\left\{\mathcal{F}^{-1}\left[\sqrt{P(f)}\mathcal{F}\left[\delta\xi(t)\right]\right]\right\}.
\end{equation}

This numerical projection guarantees that the synthesized turbulent components are mathematically confined to the real topological space $\mathbb{R}^{N}$ prior to subsequent nonlinear geometric vector compositions.

While the spectral coloring operator dictates the scale-dependent energy partition (i.e., the spectral slopes), the absolute amplitude of the fluctuations requires explicit normalization to satisfy the macroscopic turbulent energy constraints of the system. The empirical variance of the unnormalized stochastic signal $\delta\tilde{b}(t)$ obtained immediately following the inverse Fourier transform is analytically defined by the second central moment $\sigma_{emp}^{2}=\langle\delta\tilde{b}^{2}\rangle-\langle\delta\tilde{b}\rangle^{2}$.

To strictly enforce the prescribed physical variance $\sigma^{2}_{presc}$, which parameterizes the total turbulent kinetic or magnetic energy strictly associated with the specified solar wind regime, a deterministic scaling operator $\Lambda$ is applied to the time series. The finalized, energy-preserving fluctuation field $\delta b(t)$ is mathematically formalized as an affine transformation:

\begin{equation}
	\delta b(t)=\Lambda[\delta\tilde{b}(t)]=\delta\tilde{b}(t)\frac{\sigma_{presc}}{\sigma_{emp}}.
\end{equation}

By virtue of Parseval's theorem, this time-domain normalization guarantees that the discrete integral of the power spectral density over the available Nyquist frequency interval perfectly matches the physical turbulent energy prescribed by the MHD background model:

\begin{equation}
	\sum_{k=1}^{N/2}\left\vert{}\mathcal{F}\left[\delta b(t)\right]_{k}\right\vert{}^{2}\Delta f=\sigma^{2}_{presc},
\end{equation}

where $\Delta f=1/(N\Delta t)$ denotes the fundamental frequency resolution of the discrete spectrum. This necessary closure ensures that the synthesized power spectrum is not merely a scale-free fractional power law, but an exact quantitative representation of the local turbulent energy tensor.

As a whole, this procedure guarantees a properly scaled, structurally intermittent, and spectrally rigorous turbulent field.

\subsection{Covariance partitioning and cross-helicity}
\label{subsec:alfvenicity}
In MHD turbulence, the fluctuating energy tensor is intrinsically anisotropic with respect to the macroscopic guiding magnetic field $\mathbf{B}$. The proposed numerical framework enables the precise regulation of this anisotropy via the magnetic compressibility parameter, $C_{B}\in[0,1]$, formally defined as the projection of the magnetic variance along the mean field vector relative to the trace of the total covariance matrix \citep{2013LRSP...10....2B}.

Given the trace of the magnetic fluctuation covariance matrix $\sigma^{2}_{B}$, the eigenvalues of the fluctuation covariance matrix are partitioned into a parallel mode ($\delta B_{\parallel}$) and two orthogonal perpendicular modes ($\delta B_{\perp1},\delta B_{\perp2}$):

\begin{equation}
	\begin{aligned}
		& \sigma^{2}_{B,\parallel}=C_{B}\sigma^{2}_{B}\\
		& \sigma^{2}_{B,\perp1}=\sigma^{2}_{B,\perp2}=\frac{1-C_{B}}{2}\sigma^{2}_{B}.
	\end{aligned}
\end{equation}

This formulation mathematically discriminates between fast solar wind regimes, which are dynamically dominated by incompressible Alfv\'enic (perpendicular) modes ($C_{B}\to0$), and slower, collisionless regimes exhibiting strong compressive magnetosonic properties ($C_{B}\to1$). The synthesized modes are then rotated from the local field-aligned frame into the heliocentric Cartesian frame utilizing the Parker spiral geometry.

Plasma velocity fluctuations ($\delta\mathbf{v}$) are subsequently computed by mapping the magnetic fluctuations through the theoretical Alfv\'en ratio. The normalized cross-helicity parameter $r_{A}\in[0,1]$ dictates the linear correlation between the velocity and magnetic fields, effectively parameterizing the imbalance of the Els\"asser variables $\mathbf{z}^{\pm}=\delta\mathbf{v}\pm\delta\mathbf{b}/\sqrt{\mu_{0}\rho}$, where $\mu_{0}$ is the vacuum magnetic permeability. The velocity fluctuations are constructed as a linear combination of the shaped magnetic fluctuations $\delta\mathbf{b}$ and an independent stochastic compressive field $\delta\mathbf{w}$, scaled by their respective total standard deviations:

\begin{equation}
	\delta\mathbf{v}(t)=\frac{\sigma_{V}}{\sigma_{B}}\left(r_{A}\delta\mathbf{b}(t)+\sqrt{1-r_{A}^{2}}\delta\mathbf{w}(t)\right),
\end{equation}

\noindent where $sigma_{V}$ is square root of the trace of the velocity fluctuation covariance matrix $\sigma^{2}_{V}$. Imposing $r_{A}\to1$ correctly enforces the strict alignment typical of pure outward-propagating Alfv\'en wave packets, whereas setting $r_{A}\to0$ simulates a fully balanced, zero-cross-helicity state characteristic of highly reprocessed turbulence (typical of the slow solar wind). Together, these limits provide a critical diagnostic baseline for numerical algorithms sensitive to phase coherence and cross-helicity.

Before reconstructing the complete Cartesian vector fields, the synthesized localized fluctuations must be spatially propagated from the $1$ au reference boundary to the specific heliocentric coordinate of the observational node. In a continuously expanding magnetized fluid, the amplitude of turbulent perturbations does not remain spatially invariant. To ensure dynamic consistency, an anisotropic radial scaling is enforced on the fluctuating fields, strictly analogous to the Wentzel-Kramers-Brillouin (WKB) approximation for wave propagation in an inhomogeneous medium.

Specifically, the scaling of the magnetic fluctuation tensor components is intrinsically coupled to the topology of the macroscopic Parker spiral. The compressive parallel fluctuation decays analogously to the radial background field due to flux expansion, whereas the transverse Alfv\'enic fluctuations decay at a slower rate, matching the azimuthal background field behavior. This mathematically translates to the following anisotropic spatial transformations:

\begin{equation}
	\delta b_{\parallel}(r)=\delta b_{\parallel,0}\left(\frac{r_{0}}{r}\right)^{2}
\end{equation}
\begin{equation}
	\delta b_{\perp1,2}(r)=\delta b_{\perp1,2,0}\left(\frac{r_{0}}{r}\right),
\end{equation}

\noindent where the subscript $0$ denotes the stochastically generated fluctuations at the reference boundary $r_{0}$. Similarly, the thermodynamic scalar perturbations are constrained by the macroscopic expansion laws governing mass continuity and polytropic cooling, yielding:

\begin{equation}
	\begin{aligned}
		& \delta\rho(r)=\delta\rho_{0}\left(\frac{r_{0}}{r}\right)^{2}\\
		& \delta T(r)=\delta T_{0}\left(\frac{r_{0}}{r}\right)^{0.8}.
	\end{aligned}
\end{equation}

To explicitly close the fluctuation system, the spatial scaling of the velocity fluctuations is derived from their dynamic coupling with the magnetic and density fields via the Alfv\'en ratio ($\delta v\propto\delta b/\sqrt{\rho}$). Because the macroscopic mass density decays as $r^{-2}$, its square root scales as $r^{-1}$. Consequently, the transverse velocity fluctuations mathematically preserve their absolute amplitude ($\delta v_{\perp1,2}\propto r^{-1}/r^{-1}=r^{0}$), whereas the compressive parallel velocity fluctuations strictly decay inversely with the heliocentric distance ($\delta v_{\parallel}\propto r^{-2}/r^{-1}=r^{-1}$). This imposes the following anisotropic spatial transformations for the velocity field:

\begin{equation}
	\begin{aligned}
		& \delta v_{\parallel}(r)=\delta v_{\parallel,0}\left(\frac{r_{0}}{r}\right)\\
		& \delta v_{\perp1,2}(r)=\delta v_{\perp1,2,0}.
	\end{aligned}
\end{equation}

This anisotropic deformation of the fluctuation tensor guarantees that the spectral energy density and the magnetic compressibility $C_{B}$ explicitly evolve as a function of the heliocentric distance, providing a mathematically accurate boundary condition at the arbitrary multi-point detection grid.

It is mathematically pertinent to note that this specific formulation enforces a $2.5$-dimensional topological decoupling. The deterministic stationary solutions (the macroscopic Parker spiral guiding field $\mathbf{B}$ and the advective bulk flow $\mathbf{V}$) are strictly confined to the $2$D ecliptic manifold, imposing $B_{z}\equiv0$, identically. Conversely, the stochastically generated turbulent manifold is inherently fully $3$D, explicitly exciting the invariant out-of-plane orthogonal mode. This asymmetric dimensionality faithfully reproduces the physical constraints of the inner heliosphere, where macroscopic out-of-plane drifts are negligible, yet microscopic transverse Alfv\'enic perturbations are fully isotropic in the normal plane.

A critical topological feature of this numerical framework lies in the geometric projection of the synthesized fluctuation modes onto the heliocentric Cartesian frame. The local orthogonal basis $(\hat{u}_{\parallel},\hat{u}_{\perp1},\hat{u}_{\perp2})$ is not a static geometrical construct, but a dynamic, time-dependent curvilinear manifold. Because the macroscopic bulk speed $V(t)$ fluctuates across the stream interaction regions, the local Parker spiral angle $\psi(t)$ at each observational node inherently varies at each discrete temporal step:

\begin{equation}
	\psi(t)=\arctan\left(\frac{\Omega_{\odot}r_{node}}{V(t)}\right).
\end{equation}

Consequently, the rotation matrix linking the field-aligned turbulence to the ecliptic plane evolves dynamically, defining a rotating Frenet-Serret-like frame where $\hat{u}_{\parallel}(t)=(\cos\psi(t),\sin\psi(t),0)$, $\hat{u}_{\perp1}(t)=(-\sin\psi(t),\cos\psi(t),0)$, and the out-of-plane vector remains fixed as $\hat{u}_{\perp2}=(0,0,1)$.

Furthermore, from a numerical perspective, simulating highly coupled stochastic differential systems requires control over the multidimensional probability spaces. To mathematically prevent the emergence of spurious off-diagonal covariance terms (artificial cross-helicity or anomalous thermal-magnetic correlations), the Gaussian white noise innovations for each topological component (parallel, perpendicular, density, and temperature) across both the fast and slow regimes are strictly decoupled. This is achieved by initializing the Pseudo-Random Number Generator (PRNG) with uniquely spaced, deterministic seed offsets. This rigid state management guarantees that the underlying stochastic subspaces remain perfectly orthogonal, ensuring that any statistical correlation in the final synthetic fields is exclusively the analytical result of the prescribed nonlinear mapping (such as the Alfv\'enicity operator $r_{A}$), rather than numerical artifacts of the Monte Carlo generation process.

Beyond the internal decoupling of the state variables, the present framework enforces a strict mathematical condition on the spatial cross-covariance matrix of the spatially distributed multi-point network. In a highly turbulent medium, the transverse correlation length of the fluctuating fields is inherently finite. Provided that the macroscopic spatial separation between distributed observational nodes significantly exceeds this characteristic correlation scale, the individual nodes effectively sample distinct, mutually uncorrelated turbulent flux tubes.

To numerically formalize this macroscopic decorrelation for widely distributed spatial nodes, the stochastic cascades are not generated globally and subsequently advected; rather, they are synthesized locally at each node. The underlying probability space for the $i$-th spatial node is strictly segregated by initializing its PRNG branch with a deterministic, non-overlapping stride offset:

\begin{equation}
	\mathcal{S}_{i}=\mathcal{S}_{master}+i\cdot\Delta_{seed},
\end{equation}

\noindent where $\mathcal{S}_{master}$ is the global initial state and $\Delta_{seed}$ ensures orthogonal memory allocation. Consequently, while the deterministic macroscopic dynamics (e.g., advected stream interfaces) yield a strong cross-correlation across the discrete network topology, the superimposed zero-mean stochastic fluctuations remain mutually orthogonal between any two distinct nodes $i$ and $j$ at distances larger than the correlation length:

\begin{equation}
	\langle\delta\mathbf{b}_{i}(t)\cdot\delta\mathbf{b}_{j}(t+\tau)\rangle=0\quad\forall\tau,\quad\text{if }i\neq j.
\end{equation}

This spatial decorrelation prevents artificial phase coherence in the multipoint analysis, ensuring that the validation of distributed network algorithms is challenged by physically realistic, statistically independent noise realizations at each localized sampling point.

From a kinematic perspective, the mathematical formulation enforces a strict dimensionality reduction on the macroscopic flow, while preserving the full degrees of freedom of the turbulence. The deterministic bulk solar wind is modeled as a purely $1$D radial expansion flow ($V\hat{e}_{x}$). Conversely, the synthesized stochastic velocity fluctuations $\delta\mathbf{v}$ are fully $3$D. To project these fluctuations from the local magnetic frame back to the heliocentric Cartesian system, a time-dependent geometric transformation is applied:

\begin{equation}
	\begin{aligned}
		& V_{x}(t)=V(t)+\delta v_{\parallel}(t)\cos\psi(t)-\delta v_{\perp1}(t)\sin\psi(t)\\
		& V_{y}(t)=\delta v_{\parallel}(t)\sin\psi(t)+\delta v_{\perp1}(t)\cos\psi(t)\\
		& V_{z}(t)=\delta v_{\perp2}(t).
	\end{aligned}
\end{equation}

This linear superposition explicitly couples the transverse and longitudinal fluctuations to the macroscopic radial flow $\mathbf{V}(t)$ via the dynamical Parker angle $\psi(t)$.

Conversely, the magnetic field reconstruction mandates a rigorous nonlinear vector composition to ensure the exact mathematical decoupling of compressive and directional fluctuations. Specifically, an intermediate strictly transverse field $\tilde{\mathbf{B}}(t)$ is first generated by adding the perpendicular fluctuations to the mean guiding field $\mathbf{B}$:

\begin{equation}
	\tilde{\mathbf{B}}(t)=\mathbf{B}+\delta b_{\perp1}(t)\hat{u}_{\perp1}(t)+\delta b_{\perp2}(t)\hat{u}_{\perp2}.
\end{equation}

Simultaneously, a prescribed scalar magnitude is analytically defined to precisely encompass the background intensity $\vert{}\mathbf{B}\vert{}$ and the parallel compressive perturbation:

\begin{equation}
	\vert{}\mathbf{B}_{presc}(t)\vert{}=\vert{}\mathbf{B}\vert{}+\delta b_{\parallel}(t).
\end{equation}

The finalized $3$D magnetic field vector $\mathbf{B}(t)$ is subsequently computed through a nonlinear, norm-preserving projection operator that geometrically scales the directional vector to match the exact prescribed compressive magnitude:

\begin{equation}
	\mathbf{B}(t)=\tilde{\mathbf{B}}(t)\frac{\vert{}\mathbf{B}_{presc}(t)\vert{}}{\vert{}\tilde{\mathbf{B}}(t)\vert{}}.
\end{equation}

This specific nonlinear mapping is of paramount importance: it ensures that the injected parallel cascade explicitly and exclusively models the true physical compressibility of the medium (i.e., true physical fluctuations in the magnetic modulus), strictly avoiding spurious compressive artifacts that would inevitably arise from a na\"ive linear Cartesian superposition of finite-amplitude orthogonal perturbations.

Finally, the reconstruction of the localized thermodynamic scalar fields (namely, the plasma mass density $\rho(t)$ and the macroscopic proton temperature $T(t)$) is achieved by superimposing the independent stochastic compressive fluctuations onto the dynamically mixed macroscopic profiles. However, since the turbulent perturbations are stochastically driven, large-amplitude negative deviations could theoretically violate local thermodynamic constraints (e.g., yielding unphysical negative densities or temperatures). To mathematically guarantee strict positivity across the entire domain, a nonlinear thresholding operator (clipping) is enforced element-wise on the resultant scalar fields:

\begin{equation}
	\rho(t)=\max\left(\rho_{min},\rho(t)+\delta\rho(t)\right)
\end{equation}

\noindent and

\begin{equation}
	T(t)=\max\left(T_{min},T(t)+\delta T(t)\right),
\end{equation}

\noindent where $\rho_{min}$ and $T_{min}$ are predefined absolute lower bounds, ensuring that the synthesized environment remains strictly physical and numerically stable for all subsequent derivations, such as the local Alfv\'en speed or the plasma $\beta$ parameter (defined as the ratio of the fluid thermal pressure to the magnetic pressure).

\section{Analytical modeling of nonlinear heliospheric structures and turbulence modulation}
\label{sec:transients}
While the stochastic solar wind background formalized in \S{} \ref{sec:synthetic_environment} represents the quiescent continuous state of the heliosphere, the true operational challenge for distributed multi-point observatories lies in the detection and geometric resolution of transient, highly energetic macroscopic structures. The proposed framework deterministically perturbs the synthetic turbulent environment by injecting exact analytical solutions for CMEs, CIRs, and HCS crossings. This section details the mathematical physics governing the macroscopic topology and kinematic evolution of these nonlinear phenomena.

\subsection{Coronal Mass Ejections: force-free topologies and sheath dynamics}
\label{subsec:cme_modeling}
The core of a CME, frequently observed in-situ as a Magnetic Cloud (MC), exhibits a highly organized, magnetically closed internal topology. This structure is typically characterized by strongly enhanced magnetic field strength, a smooth, large-scale coherent rotation of the field vector, and an anomalously low proton temperature indicative of its isolated coronal origins \citep{1981JGR....86.6673B}. To mathematically reproduce this complex $3$D topology within a continuous Eulerian grid, the MC is formulated as a macroscopic, self-contained magnetic flux rope.

Assuming that the plasma $\beta$ parameter inside the cloud is intrinsically low ($\beta\ll1$), the internal magnetic pressure overwhelmingly dominates the thermal kinetic pressure of the plasma. Under these specific thermodynamic conditions, the internal magnetic configuration naturally relaxes toward a minimum-energy state. Consequently, the flux rope is rigorously modeled as a force-free field, satisfying the nonlinear differential constraint where the current density inside the MC ($\mathbf{j}_{MC}=\nabla\times\mathbf{B}_{MC}$) is strictly strictly parallel to the MC magnetic field $\mathbf{B}_{MC}$:

\begin{equation}
	\nabla\times\mathbf{B}_{MC}=\alpha(\mathbf{r})\mathbf{B}_{MC},
\end{equation}

\noindent where $\alpha(\mathbf{r})$ is a scalar function dictating the spatial distribution of the magnetic helicity, which measures the degree of the magnetic field line twisting. In the present numerical model, the flux rope is formalized using the classic Lundquist analytical solution for a Linear Force-Free Field (LFFF). This formulation assumes a constant $\alpha$, which physically represents a fully relaxed Taylor state \citep{1951PhRv...83..307L}. In a local cylindrical coordinate system $(\hat{e}_{\rho},\hat{e}_{\phi},\hat{e}_{z})$ co-moving with the transient and aligned with its central longitudinal axis, the exact mathematical solutions for the MC magnetic field components are described by zeroth- and first-order ordinary Bessel functions of the first kind ($J_{0},J_{1}$):

\begin{equation}
	\begin{aligned}
		& B_{MC,\tilde{r}}(\tilde{r})=0\\
		& B_{MC,\phi}(\tilde{r})=\mathcal{H}B_{MC,0}J_{1}(\chi_{0,1}\tilde{r})\\
		& B_{MC,z}(\tilde{r})=B_{MC,0}J_{0}(\chi_{0,1}\tilde{r}),
	\end{aligned}
\end{equation}

\noindent where $B_{MC,0}$ defines the prescribed magnetic field intensity at the rope's central axis, and $\mathcal{H}=\pm1$ represents the strict binary macroscopic magnetic helicity (defining right-handed or left-handed chirality, respectively). The dimensionless radial coordinate is defined as $\tilde{r}=r/R_{MC,0}\in[0,1]$, where $R_{MC,0}$ is the macroscopic radius of the flux rope boundary. To satisfy the physical boundary condition where the axial field strictly vanishes at the cloud's external edge ($B_{MC,z}(1)=0$), the constant scaling coefficient is anchored to the first fundamental root of the zeroth-order Bessel function, i.e., $\chi_{0,1}\approx2.4048$.

The synthetic temporal sampling of the MC signature by an arbitrary spatial observer essentially constitutes a one-dimensional Eulerian temporal sampling through this dynamic $3$-dimensional manifold. The arbitrary geometrical orientation of the flux rope within the macroscopic heliospheric Cartesian domain $(X,Y,Z)$ is mathematically parameterized by a rotational transformation matrix $\mathcal{R}_{MC}(\theta,\varphi)$. Here, $\theta\in[-\pi/2,\pi/2]$ defines the out-of-ecliptic tilt angle (governing the inclination of the transient), and $\varphi\in[0,2\pi)$ defines the azimuthal longitude. This coordinate transformation strictly dictates which specific Cartesian components will exhibit the classic unipolar core signature versus the rotating bipolar signature. The geometric intersection between the advecting flux rope and the stationary detection node is unequivocally determined by the impact parameter $p_{MC,0}\in[0,R_{MC,0}]$, which quantifies the minimum orthogonal distance between a spacecraft's linear trajectory and the central axis of the cloud. The temporal sampling along this specific intersection chord is governed by the transient's combined advective and expansive kinematics. To emulate the self-similar radial expansion of the flux rope as it propagates outward from the Sun, a time-dependent, linearly varying macroscopic velocity field $V_{MC}(t)$ is superimposed onto the center-of-mass transit speed $V_{c}$ (generally larger than the ambient solar wind speed $V$, $V_{c}>V$):

\begin{equation}
	V_{MC}(t)=V_{c}-V_{exp}\left(\frac{t-t_{c}}{\Delta t_{MC}/2}\right),
\end{equation}

\noindent where $V_{exp}$ dictates the isotropic radial expansion rate, $t_{c}$ is the transit time of the rope's center, and $\Delta t_{MC}$ is the prescribed temporal duration of the crossing.

Crucially, the explicit temporal dependence of the advection velocity $V_{MC}(t)$ introduces a kinematic asymmetry into the Eulerian temporal sampling of the flux rope. While the spatial Lundquist model is inherently symmetric with respect to its central axis ($\tilde{r}=0$), the discrete time-series mapping is governed by the nonlinear differential relation $d\tilde{r}=-[V_{MC}(t)/R_{MC,0}(t)]dt$. Because the leading edge of the physically expanding transient overtakes the stationary observer at a significantly higher relative velocity than the trailing edge, the resulting temporal magnetic profile is naturally compressed at the front and extended at the rear, mimicking the dynamical ``pancaking'' effect.

Furthermore, to satisfy the conservation of magnetic flux within the expanding ideal MHD fluid, the prescribed central magnetic field intensity $B_{MC,0}$ cannot be treated as a static parameter. As the macroscopic radius $R_{MC,0}(t)$ expands linearly over time, the MC magnetic field amplitude is dynamically scaled according to the geometric flux conservation law:

\begin{equation}
	B_{MC,0}(t)=B_{inj}\left(\frac{R_{inj}}{R_{MC,0}(t)}\right)^{\gamma_{B}},
\end{equation}

\noindent where $B_{inj}$ and $R_{inj}$ represent the intrinsic MC central field and radius parameters at the injection boundary. The polytropic expansion index $\gamma_{B}\approx2$ dictates a predominantly transverse flux expansion of the cylindrical topology. This coupled magneto-kinematic formulation guarantees that the synthesized transient accurately reproduces the non-Gaussian, skewed magnetic field profiles invariably observed in in-situ interplanetary observations \citep[e.g.,][]{2021ApJ...919L..30S}.

Finally, the insertion of the analytical flux rope into the continuous background medium requires a mathematically rigorous topological stitching to preserve the $\mathcal{C}^{\infty}$ continuity of the macroscopic state variables and prevent non-physical infinite derivatives at the flux rope boundaries. The abrupt geometric boundaries of the standard Lundquist cylinder ($\tilde{r}=1$) are numerically softened via a continuous, nonlinear topological blending operator $\Xi:\mathbb{R}\to[0,1]$. This operator is defined as a piecewise hyperbolic tangent function acting strictly across the leading and trailing edges of the MC.

The total resolved macroscopic magnetic field $\mathbf{B}_{tot}(t)$ is thus computed as a dynamically weighted convex superposition of the surrounding turbulent medium (or background solar wind) $\mathbf{B}(t)$ and the internal force-free field $\mathbf{B}_{MC}(t)$:

\begin{equation}
	\mathbf{B}_{tot}(t)=[1-\Xi(t)]\mathbf{B}(t)+\Xi(t)\mathbf{B}_{MC}(t).
\end{equation}

Simultaneously, to physically enforce the low-$\beta$ regime strictly required for the validity of the force-free assumption, an explicit thermodynamic depression operator is applied to the background proton temperature $T$. Expanding isolated plasmoids cool adiabatically at a faster rate than the ambient solar wind. To reflect this, within the MC boundaries, the macroscopic temperature $T_{MC}$ is forced to continuously decay from the shock-heated sheath values down to an anomalously low core temperature $T_{MC}\ll T$:

\begin{equation}
	T_{tot}(t)=[1-\Xi(t)]T(t)+\Xi(t)T_{MC}.
\end{equation}

This thermodynamic blending ensures that the synthetic plasma $\beta(t)$ parameter drops by several orders of magnitude exactly concurrently with the onset of the coherent magnetic rotation, perfectly simulating the universally observed thermodynamical anomalies characteristic of interplanetary magnetic clouds \citep{1981JGR....86.6673B}.

Preceding this highly organized magnetic obstacle, fast CMEs ($V_{c}>V$) invariably drive a compressive fast-mode MHD discontinuity, resulting in the formation of a turbulent sheath region. This region spans the temporal interval between the forward shock arrival ($t_{shock}$) and the MC leading edge ($t_{LE}$). The sheath is mathematically modeled as a highly compressed, thermodynamically heated, and strongly intermittent topological boundary layer, arising from the pile-up and shock-processing of the pre-existing background solar wind.

To transition from the quiescent background to the sheath domain, a nonlinear step-like compression operator, strictly analogous to the Rankine-Hugoniot jump conditions, is applied to the background thermodynamic fields ($\rho,T$) and to the transverse magnetic components. Furthermore, the sheath region is defined by a massive amplification of the turbulent energy cascade. The ambient variance $\sigma^{2}_{presc}$ of the stochastic generator is dynamically scaled to the sheath variance $\sigma^{2}_{sheath}$ by a nonlinear amplification factor $\Gamma_{sheath}\gg1$:

\begin{equation}
	\sigma^{2}_{sheath}(t)=\Gamma_{sheath}\sigma^{2}_{presc}\quad\text{for }t\in[t_{shock},t_{LE}].
\end{equation}

Coupled with the multifractal p-model formalized in \S{} \ref{sec:synthetic_environment}, this deterministic variance amplification successfully triggers the generation of highly singular, large-amplitude intermittent magnetic coherent structures. This physically accurate modeling of the sheath ensures that automated detection algorithms are rigorously tested against the most chaotic, noisy, and complex region of the CME structure, preventing artificial, non-physical smoothness at the transient's leading edge.

Conversely, a fundamental macroscopic signature of an interplanetary MC is the drastic reduction of high-frequency magnetic fluctuations inside the cloud. The closed, highly coherent magnetic topology of the flux rope effectively screens the internal plasma from the external ambient turbulence. To mathematically reproduce this quiescent internal state, the stochastic generation pipeline must explicitly enforce a variance quenching mechanism.

Coupled with the topological blending operator $\Xi(t)$, the prescribed variance of the turbulent cascade $\sigma^{2}_{presc}$ is actively suppressed within the MC boundaries to $\sigma^{2}_{MC}<\sigma^{2}_{presc}$. This is formalized by introducing a dimensionless quenching factor $\eta_{quench}\ll1$:

\begin{equation}
	\sigma^{2}_{MC}(t)=\eta_{quench}\sigma^{2}_{presc}.
\end{equation}

Consequently, as a synthetic observer traverses the leading edge of the transient, the massive stochastic fluctuations characteristic of the shock-heated sheath ($\sigma^{2}_{sheath}$) abruptly decay into the smooth, highly phase-coherent deterministic rotation of the force-free Lundquist field ($\sigma^{2}_{MC}\ll\sigma^{2}_{sheath}$). This localized variance suppression is critical to test the robustness of automated detection algorithms, ensuring they accurately resolve the topological boundary between the noisy shock environment and the coherent magnetic obstacle without false-negative dropouts.

Ultimately, the definitive capability of the framework lies in its ability to simultaneously project this macroscopic $3$D transient across an arbitrary spatially distributed observational grid. The mathematical formulation must dynamically account for the distinct spatial coordinates of each $i$-th detection node within the network, defined by position vectors $\mathbf{r}_{i}=(r_{i},\theta_{i},\phi_{i})$ for $i=1,\dots,N_{node}$.

Because the CME propagates radially outward along a specific principal axis $\mathbf{n}_{CME}$, the temporal and spatial intersection parameters are uniquely governed by the geometric projection of each node relative to this propagation vector. Consequently, the global impact parameter $p_{MC,0}$ is substituted by a localized discrete set of impact parameters $p_{0,i}$:

\begin{equation}
	p_{0,i}=\vert{}\mathbf{r}_{i}\times\mathbf{n}_{CME}\vert{}.
\end{equation}

This formulation dictates that a node located near the principal propagation axis will sample the deep core of the magnetic cloud ($p_{0,i}\to0$), measuring extreme magnetic intensities and full field rotations. Conversely, a widely separated node within the network may only graze the flanks of the transient ($p_{0,i}\to R_{MC,0}$), observing heavily attenuated magnetic signatures and entirely distinct apparent chirality.

Furthermore, the arrival times of the shock front $t_{shock,i}$ and the center of mass $t_{c,i}$ at each $i$-th observer are dynamically staggered. This macroscopic phase delay is calculated by integrating the kinematic expansion of the CME front over the spatial separation between the nodes:

\begin{equation}
	t_{c,i}=t_{c}+\frac{(\mathbf{r}_{i}-\mathbf{r}_{ref})\cdot\mathbf{n}_{CME}}{V_{c}},
\end{equation}

\noindent where $t_{c}$ is the MC nominal center-of-mass arrival time at an arbitrary reference node located at $\mathbf{r}_{ref}$. By strictly enforcing these nonlinear spatio-temporal offsets and observer-specific impact parameters, the numerical simulator correctly breaks the spatial degeneracy inherent to single-point measurements. This multi-point projection generates a set of highly heterogeneous, yet physically correlated, synthetic time series, providing the essential mathematical foundation required to solve the inverse problem of $3$D CME triangulation in the downstream analytics pipeline.

A fundamental topological distinction must be mathematically formalized regarding the spatial domains of the shock front and the magnetic flux rope. While the MC is strictly confined within its macroscopic cylindrical boundary $R_{MC,0}$, the upstream compressive shock surface inherently possesses a larger transverse radius of curvature, effectively acting as a broad bow shock ahead of the transient.

Consequently, the spatial support of the sheath domain $\mathcal{D}_{sheath}$ strictly contains, but is geometrically significantly wider than, the flux rope domain $\mathcal{D}_{MC}$. In a spatially-distributed observers projection, if a peripheral node's impact parameter exceeds the flux rope radius ($p_{0,i}>R_{MC,0}$) but remains within the shock's lateral extent, the topological blending operator evaluates to zero identically ($\Xi(t)\equiv0\quad\forall t$), and the internal Lundquist field is bypassed. However, the transient compression operator $\Gamma_{sheath}$ remains fully active. This formulation correctly simulates grazing encounters or flank intersections, where a distributed node may record a fast forward shock and a highly turbulent sheath signature without ever traversing the coherent magnetic cloud. This geometric degeneracy represents exactly the class of ill-posed single-point observational challenges that sub-L$1$ multi-point architectures are specifically designed to resolve.

\subsection{Corotating Interaction Regions: kinematic steepening and spatial projections}
\label{subsec:cir_modeling}
In contrast to the episodic, radially expanding transients represented by CMEs, CIRs are macroscopic, quasi-stationary hydrodynamic structures inherently tied to the solar rotation. They manifest when a fast, tenuous solar wind stream, typically originating from a persistent coronal hole, overtakes a preceding slow, dense equatorial stream. Because the solar wind plasma is highly conductive and fundamentally collisionless, the streams cannot interpenetrate; instead, they undergo severe nonlinear steepening, forming a bounded interaction region governed by strong dynamic pressure gradients \citep{1999SSRv...89...21G}.

To mathematically represent this phenomenon within the proposed continuous Eulerian framework while strictly avoiding the numerical instabilities associated with discrete, discontinuous shock modeling (e.g., rigid Rankine-Hugoniot step functions), the macroscopic transition between the slow and fast velocity regimes is governed by the $\mathcal{C}^{\infty}$ continuous transition operator $\chi(t)$ introduced in \S{} \ref{subsec:solar_wind_background}. The macroscopic radial velocity field $V_{r}(t)$ does not exhibit discrete forward and reverse shock boundaries; instead, it is formalized as a smooth, strictly monotonic hyperbolic tangent shear layer:

\begin{equation}
	V_{r}(t)=V_{S}[1-\chi(t)]+V_{F}\chi(t),
\end{equation}

\noindent where $V_{S}$ and $V_{F}$ are the slow and fast wind speeds, respectively. By modulating the layer width parameter $w_{s}$ embedded within $\chi(t)$ (see \S{} \ref{subsec:solar_wind_background}), this continuous formulation flexibly encapsulates both the gradual velocity gradients of young SIRs (characteristic of the unsteepened inner heliosphere within $1$ au) and the highly compressed, discontinuous interfaces of fully developed CIRs (where nonlinear accumulation in the outer heliosphere, that is beyond $1$ au, effectively generates forward and reverse shocks).

The most complex mathematical challenge lies in modeling the localized dynamic pile-up of plasma (mass density $\rho$ and temperature $T$) and transverse magnetic flux at the stream interface. Rather than injecting artificial, ad-hoc localized Gaussian peaks, the framework generates the physical pile-up as an exact nonlinear consequence of the coupled fluid transitions. Specifically, a continuous and nonlinear thermodynamic amplification operator $\Gamma_{CIR}(t)$ is defined using the identical spatial transition function:

\begin{equation}
	\Gamma_{CIR}(t)=1\cdot[1-\chi(t)]+r_{c}\chi(t),
\end{equation}

\noindent where $r_{c}$ is the theoretical shock compression ratio. The macroscopic scalar fields and the transverse magnetic components are computed as the direct algebraic product of the background transition $U(t)$ and the localized compression operator. Thus, the generic scalar field $U(t)$ is amplified to $U_{CIR}(t)$ as:

\begin{equation}
	U_{CIR}(t)=U(t)\Gamma_{CIR}(t)=\Big(U_{S}[1-\chi(t)]+U_{F}\chi(t)\Big)\Big(1\cdot[1-\chi(t)]+r_{c}\chi(t)\Big).
\end{equation}

From a calculus perspective, because the background thermodynamic fields monotonically decrease across the interface (e.g., $\rho_{S}>\rho_{F}$), while the compression operator monotonically increases ($1<r_{c}$), their algebraic product intrinsically yields a highly localized, non-uniform quadratic maximum exactly centered at the stream interface ($\chi(t)\approx0.5$). This continuous formulation reproduces the massive dynamic pressure pile-up strictly required by mass conservation, without introducing any discontinuous boundaries or arbitrary peaking functions.

Regarding the microscopic turbulent cascade, the large shear stresses within the CIR do not require the injection of artificial, ad-hoc variance amplification envelopes. Instead, the localized multiscale fluctuations are dynamically constructed through the continuous, variance-preserving topological blending of the independent slow and fast multifractal cascades. As previously formalized in \S{} \ref{subsec:solar_wind_background}, this mixing is rigorously governed by the square root of the continuous transition operator, e.g., $\delta\mathbf{b}(t)=\delta\mathbf{b}_{F}(t)\sqrt{\chi(t)}+\delta\mathbf{b}_{S}(t)\sqrt{1-\chi(t)}$. This mathematical formulation guarantees that the spectral indices, the magnetic compressibility $C_{B}$, and the non-Gaussian intermittent properties of the local plasma naturally and continuously transition across the stream interface, faithfully coupling the macroscopic kinematic shear to the underlying stochastic dissipation scales.

The most critical operational utility of the developed simulator lies in its capability to project this highly anisotropic structure across the distributed $3$D spatial grid. Because the CIR represents a stationary spatial pattern in the corotating frame, its macroscopic topological boundaries form Archimedean spirals anchored to the rotating Sun. Moreover, the Stream Interface (SI) is not strictly parallel to the solar rotation axis; it exhibits a characteristic out-of-ecliptic tilt angle $\gamma_{tilt}$ tracing the warped topology of the global heliospheric magnetic field.

Consequently, the time of arrival of the SI at any generic $i$-th spatial node is uniquely dictated by a complex interplay of its radial distance $r_{i}$, heliolongitude $\phi_{i}$, and heliolatitude $\theta_{i}$. The temporal mapping of the SI arrival time $t_{SI,i}$ at the $i$-th spatial node, relative to an arbitrary reference node, is governed by the $3$D exact differential rotation equation:

\begin{equation}
	t_{SI,i}=t_{SI,ref}+\int_{r_{ref}}^{r_{i}}\frac{dr}{V_{r,SI}(r)}+\frac{\phi_{ref}-\phi_{i}}{\Omega_{\odot}}+\frac{\tan(\gamma_{tilt})(\theta_{ref}-\theta_{i})}{\Omega_{\odot}},
\end{equation}

\noindent where $t_{SI,ref}$ denotes the nominal SI arrival time at an arbitrary reference node located at the heliocentric spatial coordinates $(r_{ref},\theta_{ref},\phi_{ref})$ and $V_{r,SI}(r)$ represents the macroscopic radial advection speed of the stream interface. In a spatially distributed grid, the corotation term ($\Delta\phi/\Omega_{\odot}$) and the latitudinal tilt term ($\Delta\theta/\Omega_{\odot}$) introduce large, deterministic macroscopic phase delays, often spanning several days.

By explicitly formalizing this corotating spatial geometry, the numerical framework guarantees that synthetic time series generated for distinct longitudinal nodes correctly exhibit the invariant, time-shifted morphological signatures of the CIR. This mathematical coupling of continuous fluid-dynamic compression and Archimedean multi-point projection provides an analytical baseline to validate algorithms designed to predict the geo-effectiveness of high-speed streams based on off-axis multi-point spatial sampling.

\subsection{Heliospheric Current and Plasma Sheet crossings modeling}
\label{subsec:hcs_modeling}
Intimately coupled with the macroscopic dynamics of CIRs is the crossing of the HCS \citep{1965JGR....70.5793W}. Because fast solar wind streams typically originate from unipolar coronal holes, the fluid-dynamic transition from a slow to a fast wind regime frequently coincides with a Sector Boundary Crossing (SBC). During an SBC, the macroscopic IMF undergoes a complete topological polarity reversal, shifting from an outward to an inward sector (or vice versa) relative to the Sun.

From a topological perspective, the HCS is typically embedded within the dense, slow solar wind immediately preceding the dynamic stream interaction. Consequently, its temporal centroid is analytically anchored to the SI via a prescribed temporal offset, $t_{HCS}=t_{SI}-\Delta t_{offset}$, strictly preserving the physical phase relationship between the magnetic sector boundary and the fluid-dynamic shear layer.

To mathematically prevent non-physical infinite gradients during this macroscopic polarity reversal, the numerical model treats the HCS crossing not as a discrete step, but as a continuous topological transition governed by a localized hyperbolic blending operator. A time-dependent macroscopic polarity function $\Pi:\mathbb{R}\to[-1,1]$ is defined as:

\begin{equation}
	\Pi(t)=\tanh\left(\frac{t-t_{HCS}}{w_{HCS}}\right),
\end{equation}

\noindent where $w_{HCS}$ parameterizes the characteristic temporal thickness of the current sheet. The stationary, purely directional components of the Parker spiral guiding field (the radial $B_{r}$ and azimuthal $B_{\phi}$ components) are subsequently multiplied by this topological operator:

\begin{equation}
	\mathbf{B}_{HCS}(t)=\Pi(t)\mathbf{B}(t).
\end{equation}

This smooth zero-crossing mathematically guarantees that the invariant scalar magnitude of the background magnetic field is momentarily depressed at the exact center of the current sheet ($B_{r}=B_{\phi}=0$ at $t=t_{HCS}$), correctly simulating the local drop in magnetic pressure characteristic of the HCS core \citep{1965JGR....70.5793W}. Crucially, the topological operator $\Pi(t)$ is applied strictly to the in-plane ecliptic components. The out-of-plane orthogonal turbulent fluctuation ($\delta b_{\perp2}$) remains active and unquenched. This prevents a non-physical singularity where the total magnetic field strictly vanishes; instead, the field magnitude $\vert{}\mathbf{B}\vert{}$ drops to a finite minimum dictated solely by the local transverse turbulent noise, perfectly matching in-situ observations of current sheet cores.

Crucially, this macroscopic polarity reversal dictates a mandatory topological transformation on the turbulent cross-helicity. In the collisionless solar wind, Alfv\'enic fluctuations predominantly propagate outward from the Sun. Because the Els\"asser variables dictate that the sign of the velocity-magnetic field correlation depends explicitly on the ambient field polarity, the cross-helicity parameter $r_{A}$ must strictly invert its sign across the sector boundary to conserve the outward propagation direction of the wave packets. Consequently, the localized continuous transition operator is identically applied to the background Alfv\'enicity:

\begin{equation}
	r_{A,HCS}(t)=\Pi(t)r_{A}(t).
\end{equation}

This guarantees that the stochastically generated velocity fluctuations $\delta\mathbf{v}$ dynamically adapt their phase correlation to the reversing magnetic manifold, preventing the artificial generation of non-physical inward-propagating cascades.

Furthermore, to maintain total macroscopic pressure balance across the heliosphere, the HCS is physically embedded within the Heliospheric Plasma Sheet (HPS), a dense, cooler boundary layer exhibiting a high plasma $\beta$. To synthesize this thermodynamic anomaly, an independent, highly localized Gaussian amplification operator is superimposed strictly onto the background mass density field $\rho(t)$, instantaneously coupled with an isomorphic localized depression operator for the proton temperature $T(t)$:

\begin{equation}
	\begin{aligned}
		& \rho_{HPS}(t)=\rho(t)\left[1+\Lambda_{HPS}\exp\left(-\frac{(t-t_{HCS})^{2}}{2w_{HPS}^{2}}\right)\right]\\
		& T_{HPS}(t)=T(t)\left[1-\Delta_{HPS}\exp\left(-\frac{(t-t_{HCS})^{2}}{2w_{HPS}^{2}}\right)\right],
	\end{aligned}
\end{equation}

\noindent where $\Lambda_{HPS}$ defines the density enhancement ratio, $\Delta_{HPS}\in(0,1)$ specifies the fractional thermal depression, and $w_{HPS}>w_{HCS}$ defines the temporal scale of the plasma sheet. This mathematically represents a broad thermodynamic envelope encapsulating the sharply localized magnetic current sheet. The synchronous superposition of the HCS magnetic pressure depression and the HPS thermal pressure enhancement inherently generates a massive, localized spike in the synthetic plasma $\beta$ parameter ($\beta\gg1$), which is the definitive thermodynamic signature of sector boundaries.

By dynamically coupling the HCS topological reversal, the HPS thermodynamic anomaly, and the CIR fluid-dynamic compression into a unified mathematical framework, the presented model generates the complete, highly complex suite of macroscopic stream interactions routinely encountered by distributed multi-point heliospheric observatories.

\section{Numerical experiments and multi-point, multiscale validation}
\label{sec:validation}
To assess the predictive capabilities and the physical fidelity of the proposed unified deterministic-stochastic mathematical framework, this section presents a series of numerical experiments. The validation strategy is twofold: first, the demonstration of the accurate spatio-temporal propagation of macroscopic coherent structures across a distributed spatial domain; second, the verification that the superimposed local turbulence strictly adheres to theoretical constraints regarding spectral scaling and non-Gaussian intermittency.

\begin{figure}[ht]
	\centering
	\includegraphics[width=0.8\linewidth]{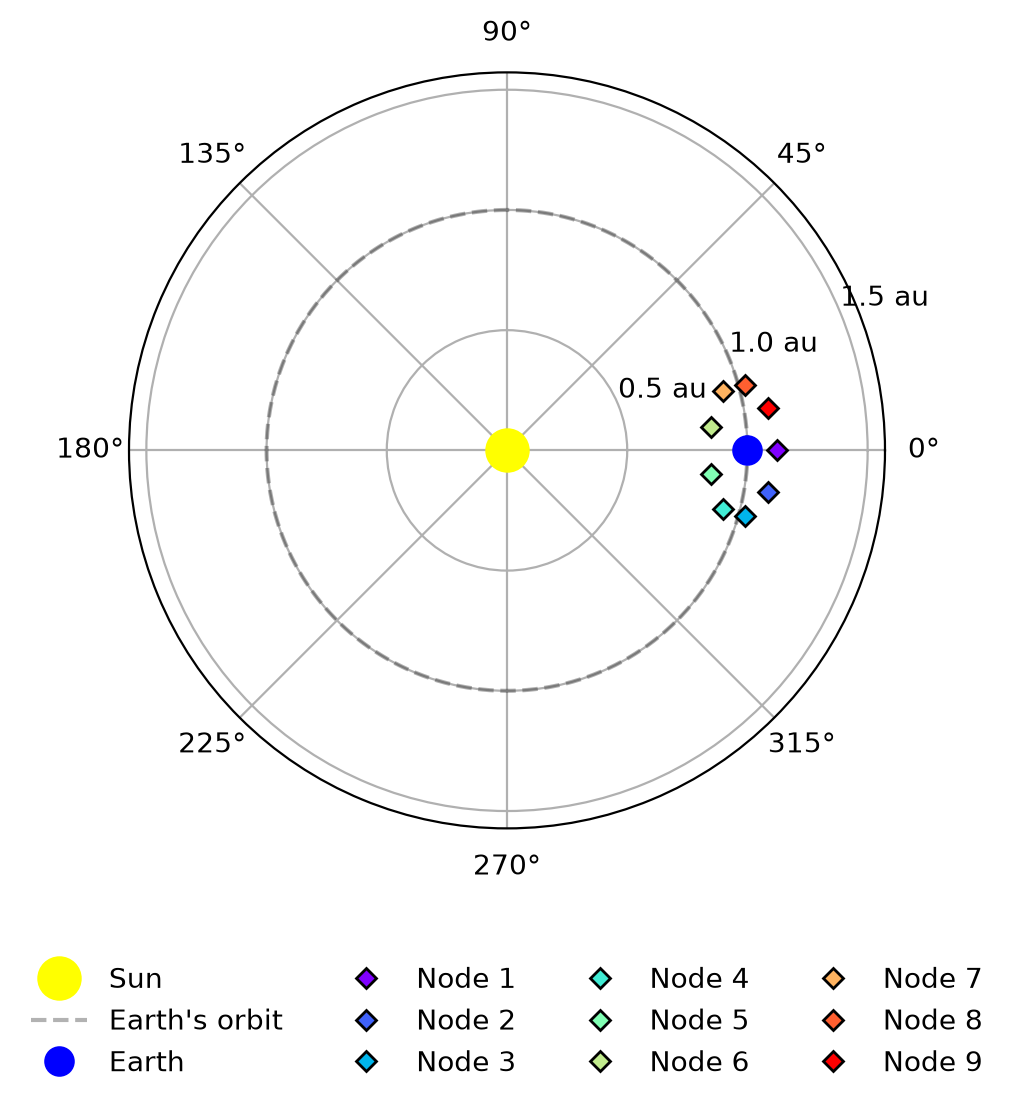}
\caption{Geometric configuration of the numerical experiment on the ecliptic plane. The origin represents the Sun (large yellow dot), while the synthetic multi-point constellation consists of nine distributed observational nodes (color-coded diamonds) deployed around the Earth ($1$ au), marked as a smaller blue dot. The Earth's orbit is displayed as dashed line. This spatially distributed array allows the direct measurement of geometric time-delays and the breaking of single-point spatial degeneracy during the transit of large-scale heliospheric structures.}
	\label{fig:constellation}
\end{figure}

For the purpose of these experiments, and to maintain strict coherence with the orbital constraints of current and near-future in-situ heliospheric missions, the synthetic observational domain is restricted to the ecliptic plane. Thus, a highly localized synthetic constellation is established, consisting of nine spatially distributed observational nodes operating concurrently. As illustrated in Figure \ref{fig:constellation}, these nodes are strategically positioned around the Earth at $1$ au, serving as an ideal testbed to capture the varying angles of incidence and the propagation delays of large-scale heliospheric transients. The exact orbital coordinates for each virtual observer are detailed in Table \ref{tab:constellation_coords}.

\begin{table}[ht]
	\centering
	\begin{tabular}{ccr}
		\hline
		\hline
		\makecell[c]{\textbf{Observational}\\\textbf{node}} &
		\makecell[c]{\textbf{Heliocentric}\\\textbf{distance [au]}} &
		\makecell[c]{\textbf{Heliographic}\\\textbf{longitude [$^{\circ}$]}}\\
		\hline
		1 & 1.12 & 0.00\\
		2 & 1.10 & 350.82\\
		3 & 1.03 & 344.62\\
		4 & 0.93 & 344.71\\
		5 & 0.85 & 353.34\\
		6 & 0.85 & 6.66\\
		7 & 0.93 & 15.29\\
		8 & 1.03 & 15.38\\
		9 & 1.10 & 9.18\\
		\hline
		\hline
	\end{tabular}
	\caption{Orbital coordinates (heliocentric distance and heliographic longitude with respect to the Sun-Earth direction) of the nine synthetic observational nodes within the ecliptic plane.}
	\label{tab:constellation_coords}
\end{table}

To resolve both the macroscopic variations and the microscopic turbulent cascade, the temporal integration at each Eulerian observational node is performed over a uniformly discretized grid of $N=8192$ points with a sampling resolution of $\Delta t=60$ seconds. This configuration yields a continuous observational window of approximately $5.68$ days, which is optimal for capturing the complete transit of large-scale structures while preserving a sufficiently extended inertial range for high-frequency spectral analysis.

\subsection{Macroscopic kinematics: Corotating Interaction Regions}
\label{subsec:experiment_cir}
The first numerical experiment tests the framework's ability to model the nonlinear kinematic steepening associated with a CIR, generated by the interaction between a slow and a fast solar wind stream. The macroscopic initial conditions for the bimodal background plasma at $1$ au, along with the theoretical shock compression ratio governing the plasma pile-up, are strictly parameterized as summarized in Table \ref{tab:cir_params}.

\begin{table}[ht]
	\centering
	\begin{tabular}{lcc}
		\hline
		\hline
		\textbf{Solar wind parameter} & \textbf{Slow wind} & \textbf{Fast wind}\\
		\hline
		Bulk speed [km s$^{-1}$] & $400$ & $800$\\
		Proton density [cm$^{-3}$] & $20$ & $3$\\
		Proton temperature [K] & $5\times10^{4}$ & $5\times10^{5}$\\[2ex]
		\multicolumn{3}{c}{\textbf{Corotating Interaction Region (CIR) parameters}}\\
		\hline
		Shock compression ratio & \multicolumn{2}{c}{$2.5$}\\
		\hline
		\hline
	\end{tabular}
	\caption{Initialization parameters for the background solar wind states at $1$ au and the macroscopic shock compression ratio governing the synthetic CIR.}
	\label{tab:cir_params}
\end{table}

\begin{figure}[ht]
	\centering
	\includegraphics[width=0.8\textwidth]{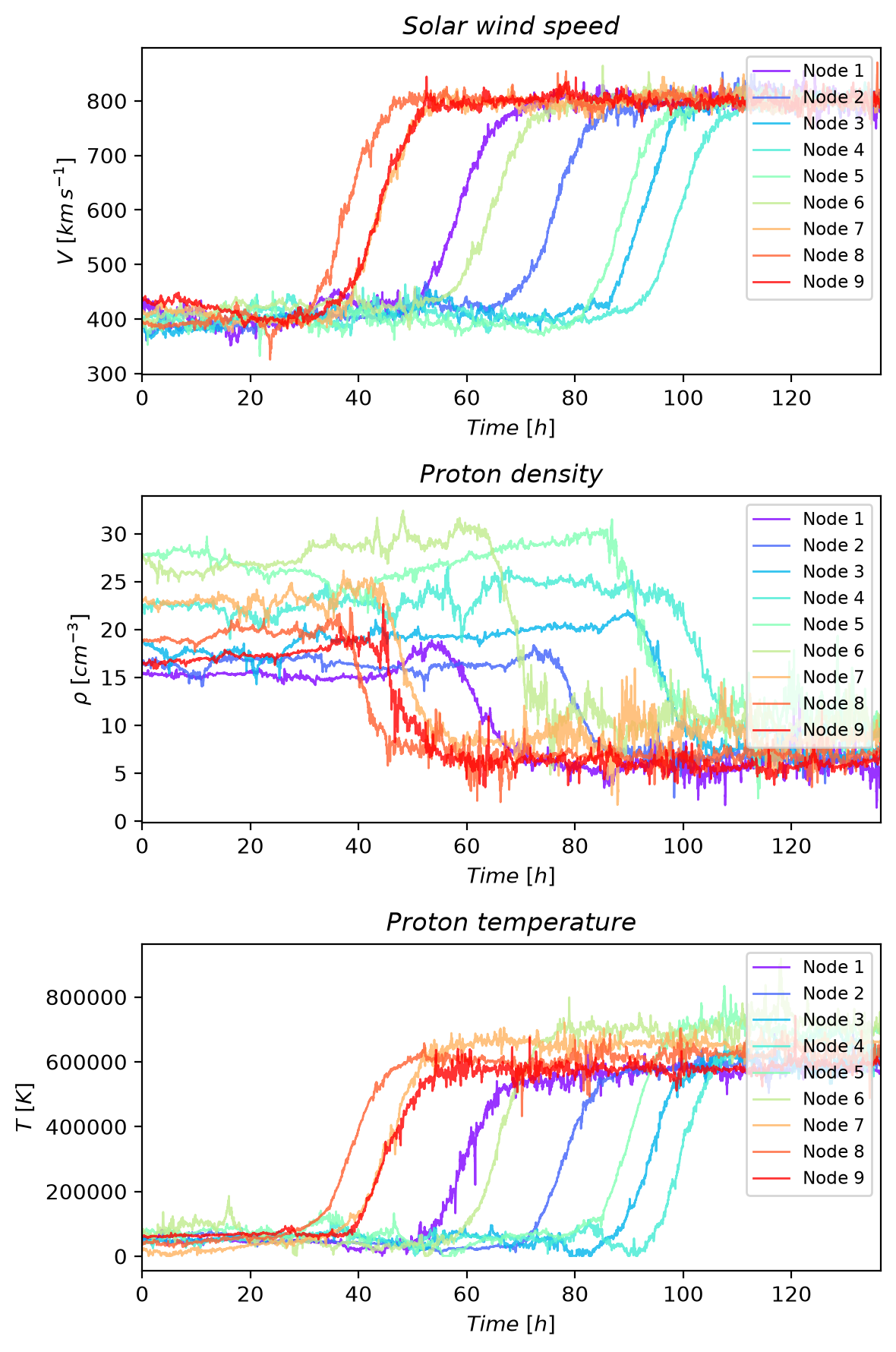}
\caption{Synthetic multi-point observations of a CIR sweeping across the nine ecliptic nodes. The three panels display, from top to bottom, the solar wind bulk speed, proton density, and temperature. The color of each temporal trace strictly corresponds to the geometric position of the respective node depicted in Figure \ref{fig:constellation}. The progressive temporal shift among the curves distinctly illustrates the tilted spatial front of the compression region.}
	\label{fig:cir_multipoint}
\end{figure}

Figure \ref{fig:cir_multipoint} displays the synthetic time series for the bulk solar wind speed, proton density, and temperature extracted simultaneously at the nine distributed nodes.

As defined in the mathematical modeling section (\S{} \ref{subsec:cir_modeling}), the stream interface is governed by a deterministic background gradient. The numerical results in Figure \ref{fig:cir_multipoint} successfully reproduce the fundamental hydrodynamic signatures of a CIR: the top panel shows the steady transition from a slow wind regime ($\approx400$ km/s) to a fast wind regime ($\approx800$ km/s). Coinciding with this velocity shear, the model dynamically triggers a localized fluid compression. This is visually evidenced by the macroscopic plasma pile-up preceding the density drop (middle panel), and the concurrent thermodynamic transition to the hotter fast-wind regime (bottom panel).

Crucially, the multi-point framework explicitly resolves the spatial degeneracy of the heliospheric structure. Because the nine nodes are distributed across different azimuthal and radial coordinates, the tilted Archimedean spiral front of the CIR intersects each node at a distinct temporal instance. This is clearly visible in the systematic phase shift of the fluid dynamic profiles across the color-coded time series. Furthermore, the varying heliocentric distances of the observers intrinsically dictate the macroscopic amplitude of the signals; for instance, the distinct vertical stratification of the density profiles strictly reflects the $r^{-2}$ radial expansion law. Together, these spatio-temporal variations provide a synthetic baseline for testing $3$D geometric triangulation algorithms based on both ToA delays and macroscopic spatial gradients. Operationally, this explicit multi-point projection is what definitively breaks the single-point spatial degeneracy: it provides the exact mathematical testbed required by mission planners to develop and train inversion algorithms capable of extracting the true $3$D propagation vector, the front's curvature, and the exact spatial extent of the co-rotating structure from discrete temporal cuts.

\subsection{Topological reversals: Heliospheric Current and Plasma Sheets crossings}
\label{subsec:experiment_hcs}
The second experiment focuses on the precise magnetic topology of the continuous background, specifically the crossing of the HCS and its associated HPS embedded within a fast solar wind stream. The macroscopic initial conditions for the background plasma at $1$ au, along with the thermodynamic parameters defining the HPS, are detailed in Table \ref{tab:hcs_params}.

\begin{table}[ht]
	\centering
	\begin{tabular}{lc}
		\hline
		\hline
		\multicolumn{2}{c}{\textbf{Fast solar wind parameters}}\\
		\hline
		Bulk speed [km s$^{-1}$] & $800$\\
		Proton density [cm$^{-3}$] & $3$\\
		Proton temperature [K] & $2.5\times10^{5}$\\[2ex]
		\multicolumn{2}{c}{\textbf{Heliospheric Plasma Sheet (HPS) parameters}}\\
		\hline
		Density enhancement [cm$^{-3}$] & $10.0$\\
		Temperature dip (fraction) & $0.5$\\
		\hline
		\hline
	\end{tabular}
	\caption{Initialization parameters for the background fast solar wind state at $1$ au and the macroscopic thermodynamic parameters defining the synthetic HPS.}
	\label{tab:hcs_params}
\end{table}

Figure \ref{fig:hcs_multipoint} presents the five-panel multipoint time series for the in-ecliptic magnetic field components, magnetic magnitude, and plasma moments.

\begin{figure}[ht]
	\centering
	\includegraphics[width=0.8\textwidth]{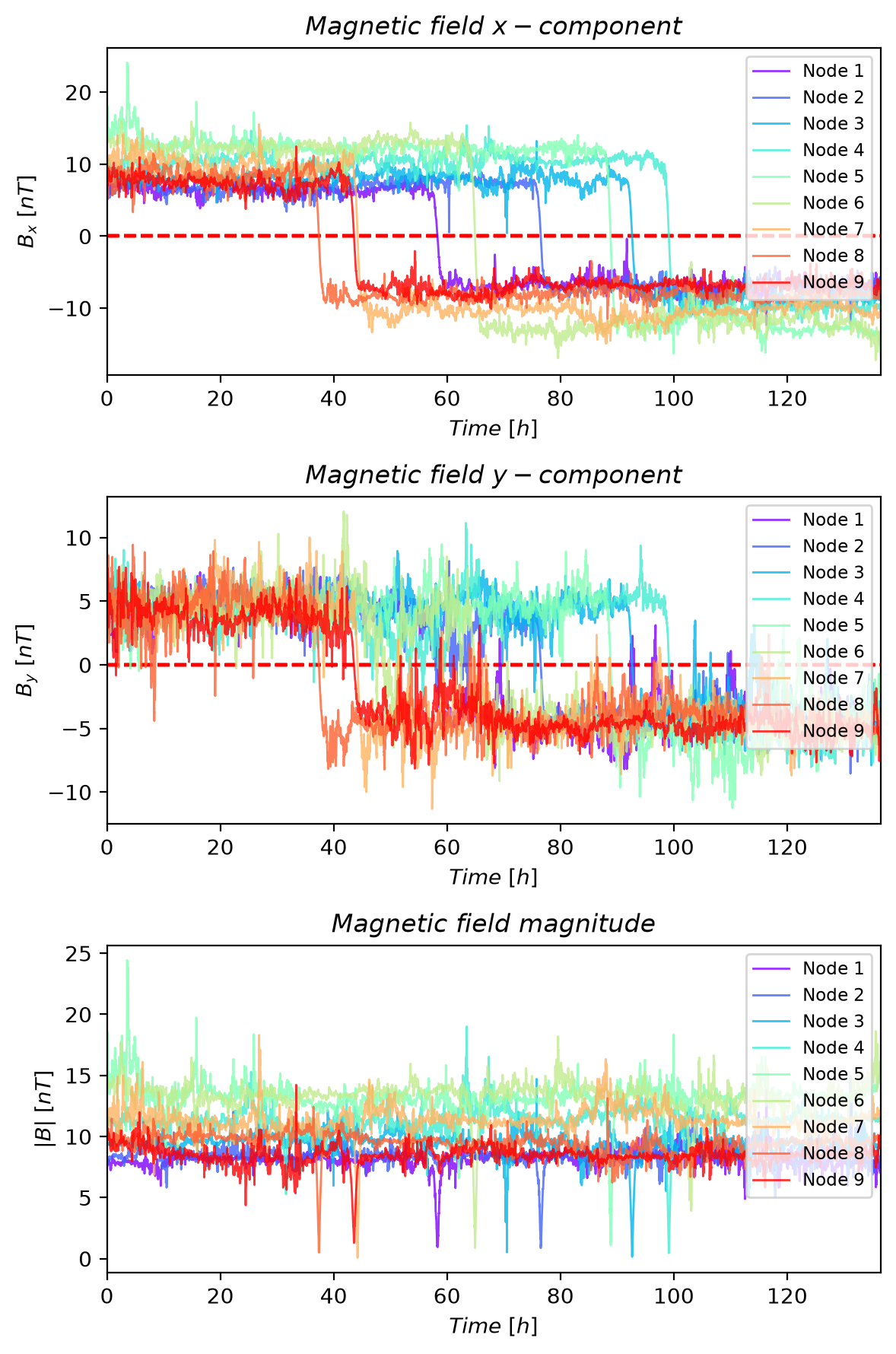}
	\caption{Multi-point synthetic time series of an HCS and HPS crossing. The five panels display the planar magnetic field components $B_{x}$ and $B_{y}$, the total magnetic field magnitude $\vert{}B\vert{}$, and the proton density and temperature. The color of each temporal trace strictly corresponds to the geometric position of the respective node depicted in Figure \ref{fig:constellation}. The topological reversal of the Parker spiral sector is perfectly captured, explicitly highlighted by the zero-crossing across the horizontal red dashed line in the first two panels, along with the localized density enhancement and temperature drop defining the HPS, with precise propagation delays across the nine observational nodes.}
	\label{fig:hcs_multipoint}
\end{figure}

\begin{figure}[ht]
	\ContinuedFloat
	\centering
	\includegraphics[width=0.8\textwidth]{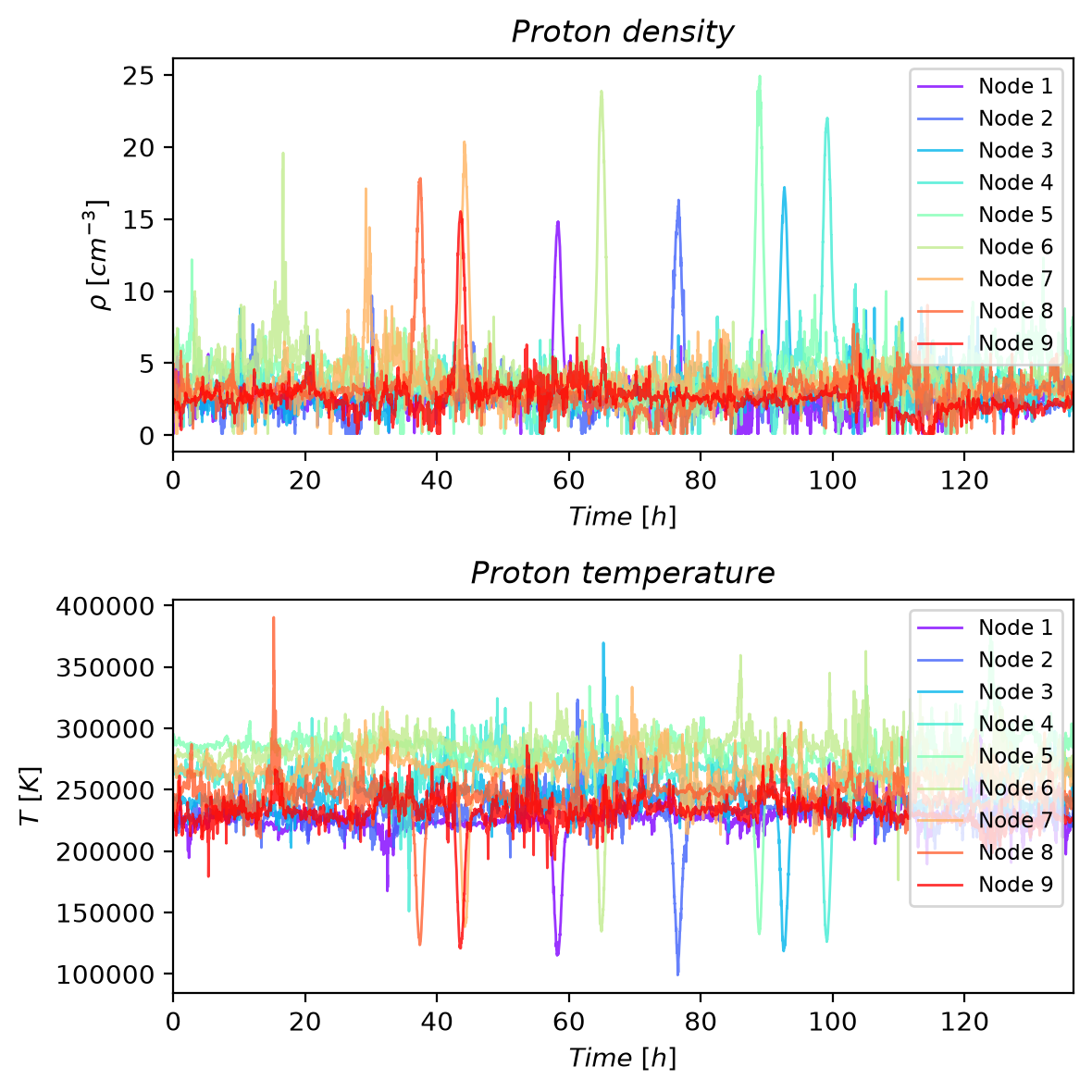}
\caption{\textit{(continued)}}
	\label{fig:hcs_multipoint}
\end{figure}

The model demonstrates absolute analytical control over the magnetic topology. As the structure sweeps over the constellation, the in-plane magnetic field components $B_{x}$ and $B_{y}$ (top two panels) exhibit a smooth yet definitive zero-crossing, perfectly mirroring the sector boundary reversal of the underlying Parker spiral. Despite the sharp topological inversion of the vector components, the scalar magnitude $\vert{}B\vert{}$ (third panel) remains continuous and physically consistent, exhibiting only a localized depression dictated by the total pressure balance constraint.

Simultaneously, the framework inherently couples this magnetic boundary with the fluid equations, generating a remarkable HPS. This is numerically validated by the pronounced, localized peak in the proton density and drop in proton temperature (bottom two panels) structurally confined within the magnetic reversal region. Once again, the differential arrival times across the nine spatially-distributed nodes highlight the model's capacity to generate spatially coherent, non-degenerate arrays of the heliospheric environment. Furthermore, the explicit radial scaling of the macroscopic fields is clearly visible in the distinct amplitude stratification of the signals, most notably in the in-plane magnetic component $B_{x}$, the proton density, and temperature, strictly reflecting their respective heliocentric distance dependencies. From an operational perspective, providing these precise distributed measurements is essential to rigorously validate data assimilation techniques. This capability enables the extraction of localized spatial gradients and drastically reduces the uncertainty in predictive models forecasting the exact terrestrial impact of geomagnetic sector boundaries.

\subsection{Mathematical metrics for stochastic turbulence and intermittency}
\label{subsec:metrics}
To systematically validate the stochastic component of the present mathematical framework, it is imperative to analyze the multiscale statistical properties of the generated synthetic time series. For this multiscale validation, the scalar and vectorial temporal arrays are extracted at a single observational node on the ecliptic plane. Three fundamental mathematical quantities are defined: the Power Spectral Density (PSD), the magnetic compressibility $C_{B}$, and the scale-dependent Flatness $\mathcal{F}(\tau)$.

The spectral energy distribution of the turbulent cascade is evaluated via the Fourier PSD, denoted as $E(f)$, which is expected to follow characteristic fractional power laws of the form $E(f)\propto f^{-\alpha}$. To quantify the degree of Alfv\'enicity versus the magnetosonic compressive nature of the fluctuations, the frequency-dependent magnetic compressibility $C_{B}(f)$ is then computed. This is rigorously defined as the ratio between the variance of the magnetic field magnitude and the total variance of the vector field fluctuations:

\begin{equation}
	C_{B}(f)=\frac{E_{\vert{}B\vert{}}(f)}{E_{B_{x}}(f)+E_{B_{y}}(f)+E_{B_{z}}(f)},
\end{equation}

where $E_{\vert{}B\vert{}}(f)$ is the PSD of the field magnitude, and the denominator represents the trace of the spectral tensor.

Finally, to assess the presence of small-scale magnetic intermittency, a hallmark of fully developed turbulence, the normalized fourth-order structure function, or flatness, $\mathcal{F}(\tau)$, is evaluated. It is defined as:

\begin{equation}
	\mathcal{F}(\tau)=\frac{\langle\Delta B(t,\tau)^{4}\rangle}{\langle\Delta B(t,\tau)^{2}\rangle^{2}},
\end{equation}

where $\Delta B(t,\tau)=B(t+\tau)-B(t)$ is the temporal increment of the field at a given time scale $\tau$, and $\langle\dots\rangle$ denotes the temporal ensemble average. For a purely Gaussian, self-similar process, $\mathcal{F}(\tau)=3$ across all scales. Conversely, a systematic increase of $\mathcal{F}(\tau)$ as $\tau\to0$ provides a rigorous mathematical proxy for the emergence of non-Gaussian, intermittent coherent structures generated by the multiplicative p-model cascade.

\subsection{The slow solar wind regime}
\label{subsec:experiment_slow_wind}
Applying the above mathematical operators, the synthetic slow solar wind regime is first investigated. The physical baseline for this numerical experiment, including both the macroscopic fluid parameters at $1$ au and the properties of the superimposed turbulent cascade, is defined in Table \ref{tab:slow_wind_params}.

\begin{table}[ht]
	\centering
	\begin{tabular}{lc}
		\hline
		\hline
		\multicolumn{2}{c}{\textbf{Slow solar wind parameters}}\\
		\hline
		Bulk speed [km s$^{-1}$] & $400$\\
		Proton density [cm$^{-3}$] & $10$\\
		Proton temperature [K] & $5\times10^{4}$\\[2ex]
		\multicolumn{2}{c}{\textbf{Turbulent properties}}\\
		\hline
		Alfv\'enicity & $0.3$\\
		Magnetic compressibility & $0.8$\\
		\hline
		\hline
	\end{tabular}
	\caption{Initialization parameters for the macroscopic fluid background and superimposed turbulence governing the synthetic slow solar wind regime at $1$ au.}
	\label{tab:slow_wind_params}
\end{table}

Figure \ref{fig:slow_wind_timeseries} illustrates the macroscopic temporal profiles at a single node for the magnetic field intensity, bulk speed, proton density, and temperature. The time series exhibit a highly variable and structurally complex environment typical of the highly reprocessed slow wind.

\begin{figure}[ht]
	\centering
	\includegraphics[width=0.8\textwidth]{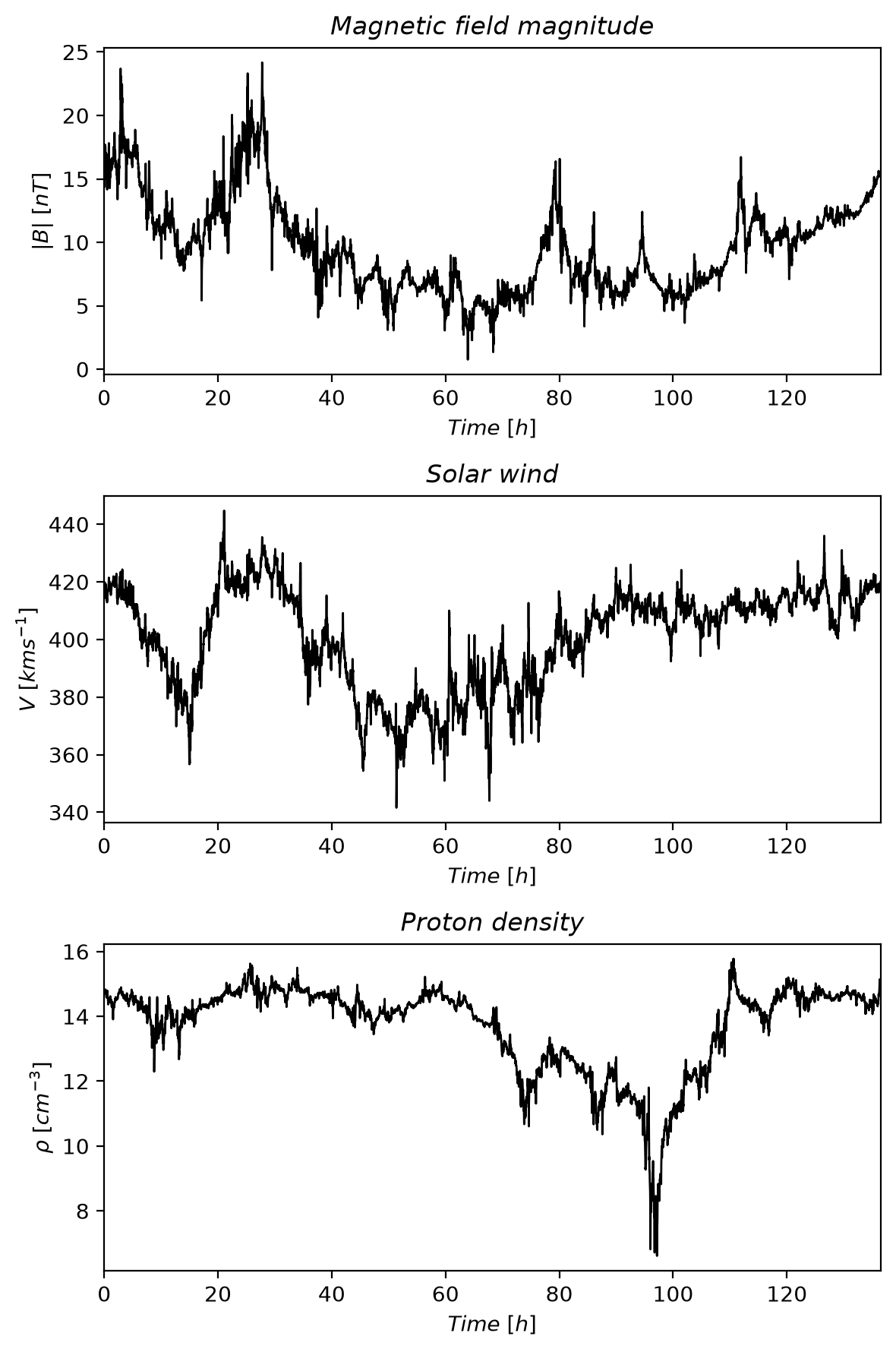}
	\caption{Macroscopic magnetic and fluid time series for the synthetic slow solar wind at a single ecliptic node. Panels from top to bottom: magnetic field magnitude, bulk speed, proton density, and temperature. The profiles correctly reflect the highly variable, compressive nature of the slow solar wind.}
	\label{fig:slow_wind_timeseries}
\end{figure}

\begin{figure}[ht]
	\ContinuedFloat
	\centering
	\includegraphics[width=0.8\textwidth]{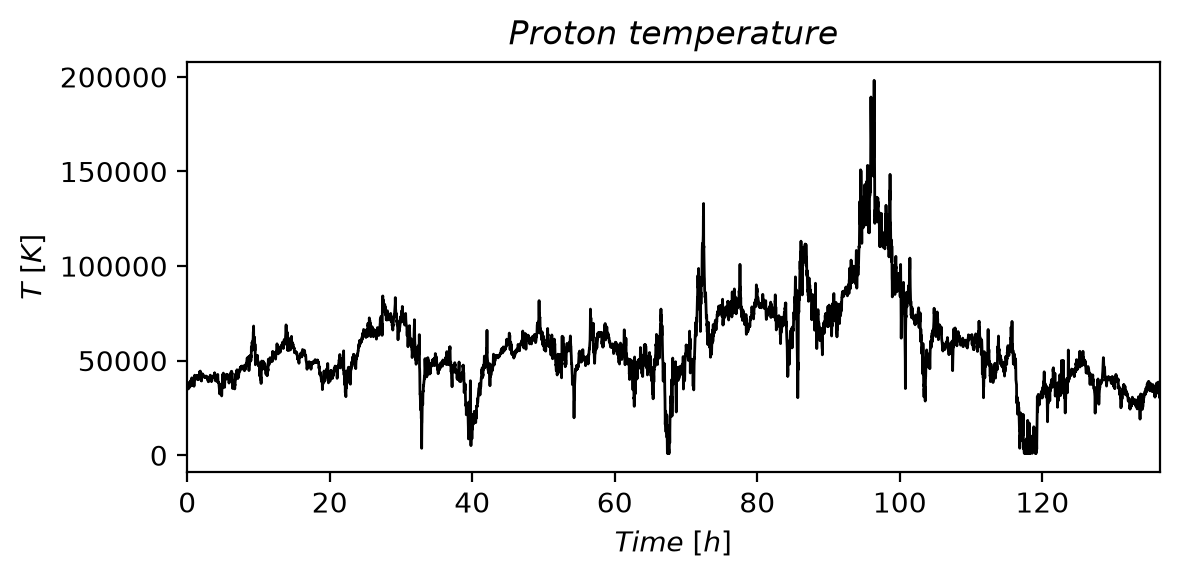}
\caption{\textit{(continued)}}
	\label{fig:slow_wind_timeseries}
\end{figure}

\begin{figure}[ht]
	\centering
	\includegraphics[width=0.8\textwidth]{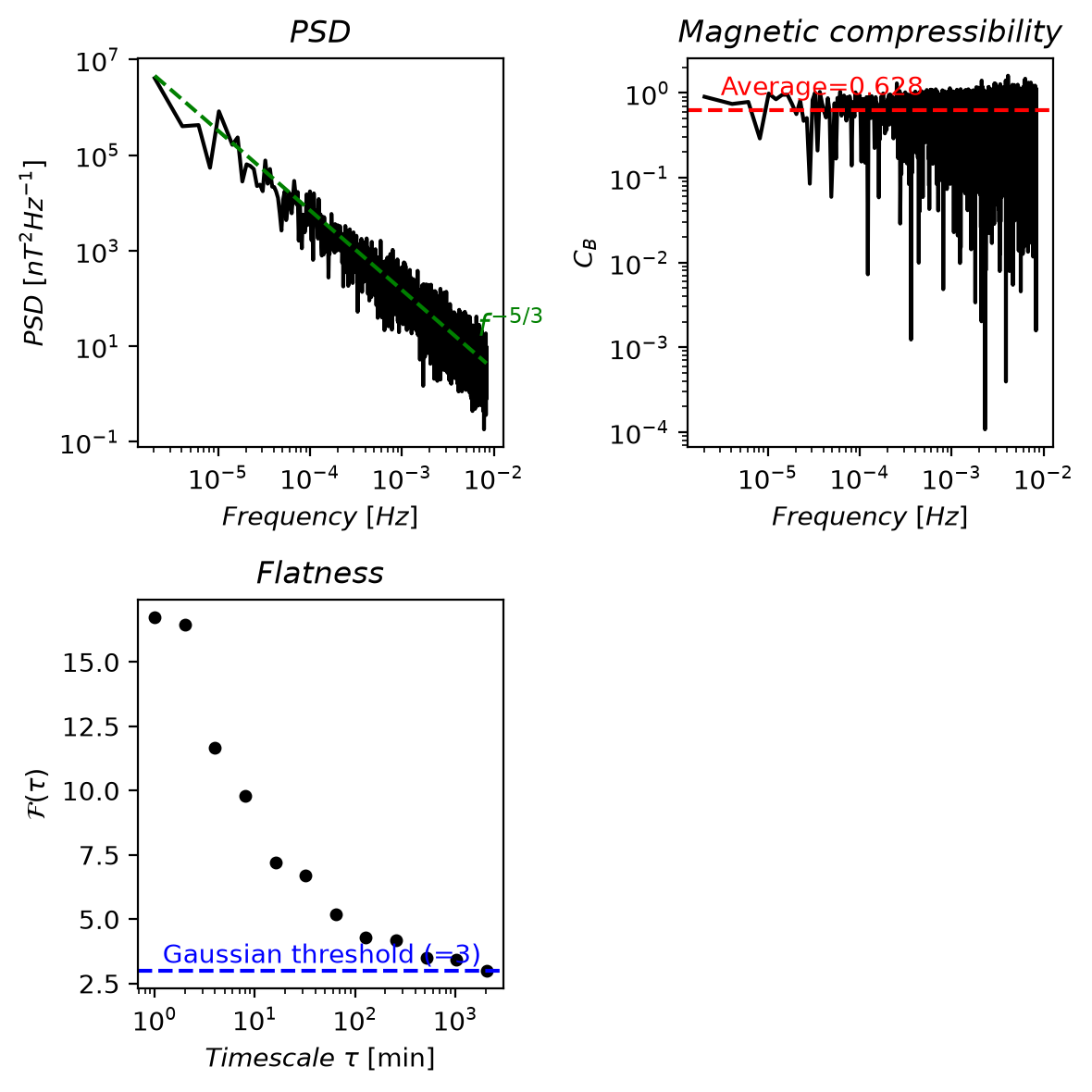}
	\caption{Multiscale statistical validation for the synthetic slow solar wind. Top-left panel shows the PSD displaying a distinct $-5/3$ spectral index, indicated by the green dashed line. Top-right panel illustrates the high magnetic compressibility $C_{B}(f)$, with an average of $0.628$ over the entire frequency range (shown as a horizontal red dashed line). Bottom panel confirms the presence of small-scale intermittency via the scale-dependent Flatness $\mathcal{F}(\tau)$, which significantly deviates from the Gaussian baseline ($mathcal{F}=3$, shown as a horizontal blue dashed line) at small scales.}
	\label{fig:slow_wind_spectra}
\end{figure}

The true physical fidelity of the model is revealed in the spectral and statistical analysis, presented in Figure \ref{fig:slow_wind_spectra}.

The PSD (top-left panel) demonstrates a fully developed turbulent cascade characterized exclusively by a Kolmogorov-like $-5/3$ spectral index over the entire resolved frequency range, perfectly aligning with theoretical expectations for the dynamically older slow wind. The magnetic compressibility $C_{B}(f)$ (top-right panel) maintains persistently high values, accurately reflecting the highly compressible and less Alfv\'enic nature of this regime. Crucially, the flatness $\mathcal{F}(\tau)$ (bottom panel) exhibits a sharp monotonic increase at small time scales ($\tau\to0$), demonstrating that the embedded p-model rigorously enforces the non-Gaussian intermittent distribution of small-scale magnetic structures.

\subsection{The fast solar wind regime}
\label{subsec:experiment_fast_wind}
The above results are contrasted by validating the framework within the fast solar wind regime, originating from coronal holes. The exact initialization parameters for this highly Alfv\'enic and less compressible environment are summarized in Table \ref{tab:fast_wind_params}.

\begin{table}[ht]
	\centering
	\begin{tabular}{lc}
		\hline
		\hline
		\multicolumn{2}{c}{\textbf{Fast solar wind parameters}}\\
		\hline
		Bulk speed [km s$^{-1}$] & $800$\\
		Proton density [cm$^{-3}$] & $3$\\
		Proton temperature [K] & $2.5\times10^{5}$\\[2ex]
		\multicolumn{2}{c}{\textbf{Turbulent properties}}\\
		\hline
		Alfv\'enicity & $0.9$\\
		Magnetic compressibility & $0.1$\\
		\hline
		\hline
	\end{tabular}
	\caption{Initialization parameters for the macroscopic fluid background and superimposed turbulence governing the synthetic fast solar wind regime at $1$ au.}
	\label{tab:fast_wind_params}
\end{table}

Figure \ref{fig:fast_wind_timeseries} presents the corresponding single-node macroscopic time series, highlighting a faster, hotter, and generally more tenuous plasma background clearly exhibiting larger Alfv\'enic fluctuations compared to the slow wind.

\begin{figure}[ht]
	\centering
	\includegraphics[width=0.8\textwidth]{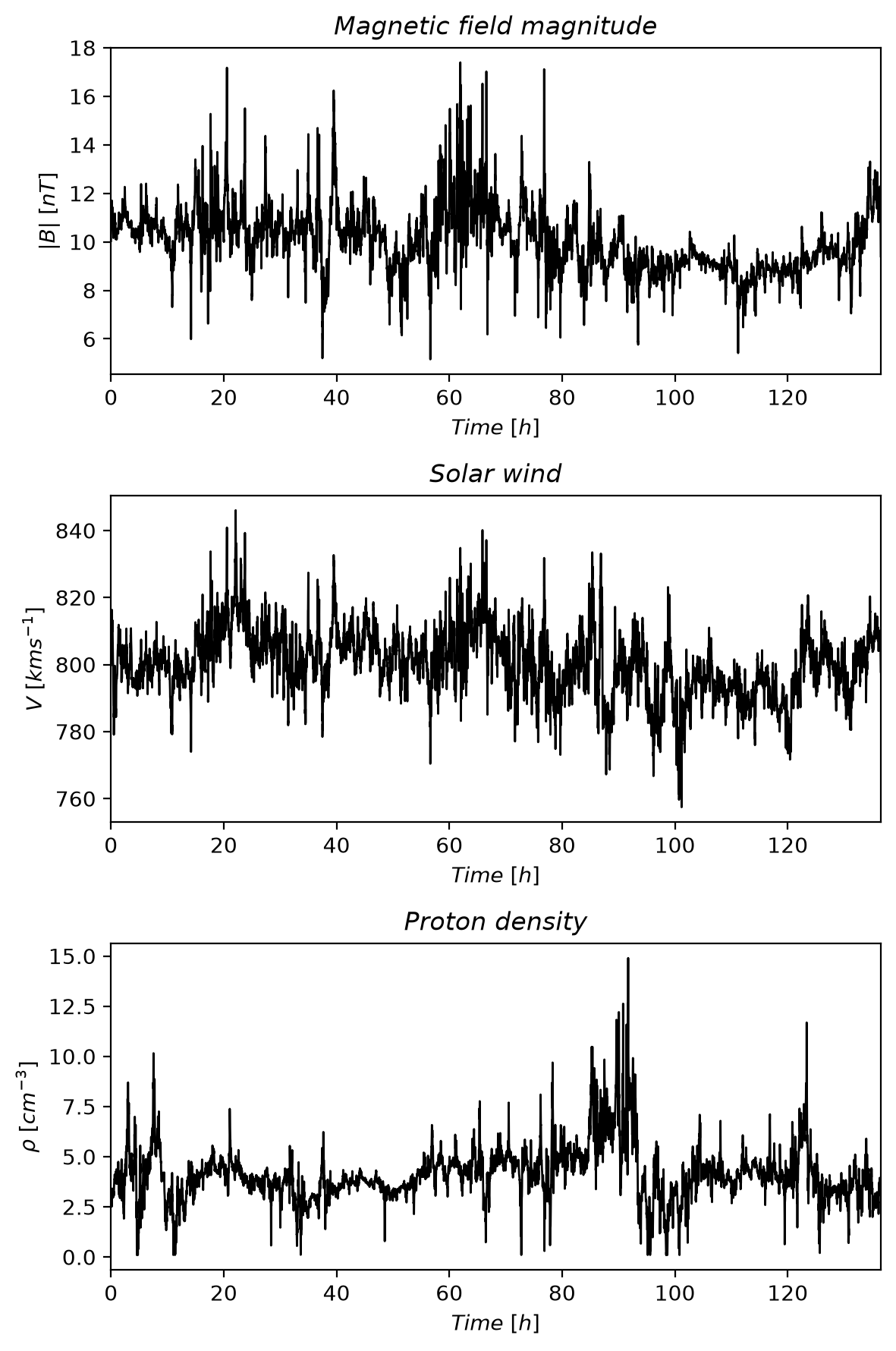}
	\caption{Macroscopic magnetic and fluid time series for the synthetic fast solar wind at a single ecliptic node. Panels from top to bottom: magnetic field magnitude, bulk speed, proton density, and temperature. The profiles correctly reflect the hotter, more tenuous, and highly Alfv\'enic nature of the fast solar wind, characterized by large-amplitude fluctuations.}
	\label{fig:fast_wind_timeseries}
\end{figure}

\begin{figure}[ht]
	\ContinuedFloat
	\centering
	\includegraphics[width=0.8\textwidth]{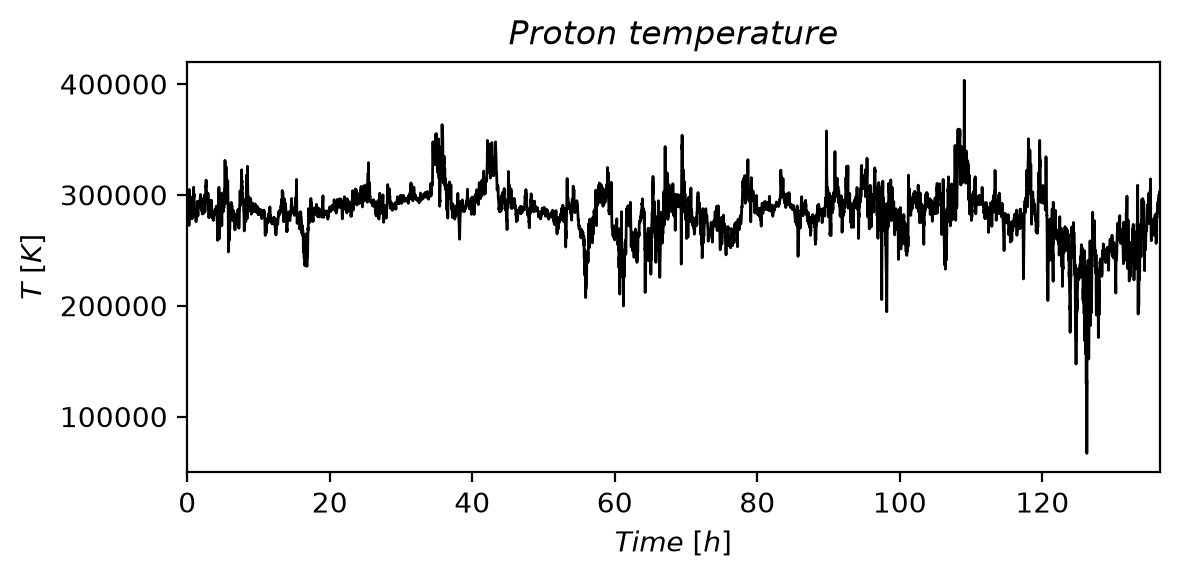}
\caption{\textit{(continued)}}
	\label{fig:fast_wind_timeseries}
\end{figure}

\begin{figure}[ht]
	\centering
	\includegraphics[width=0.8\textwidth]{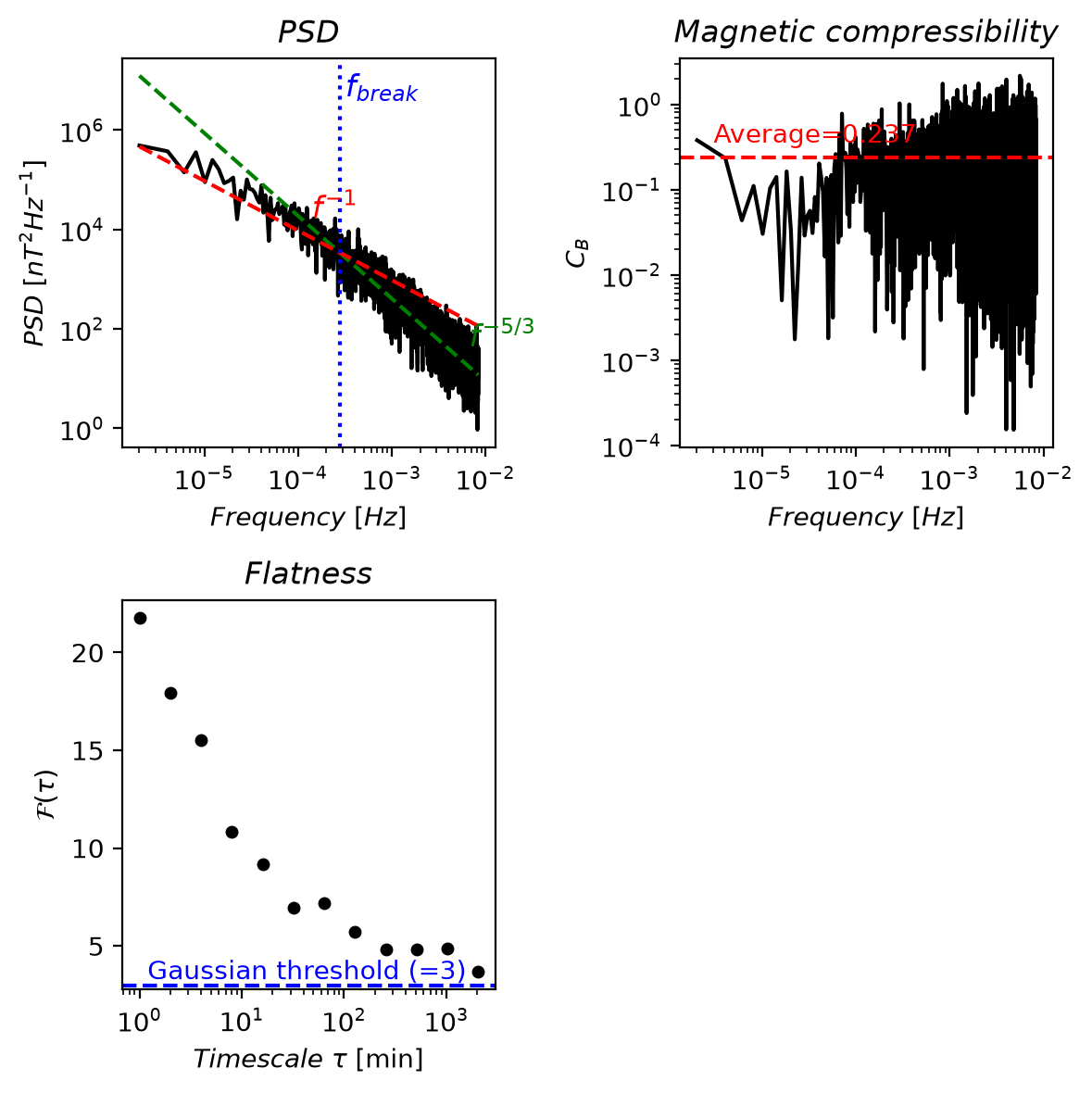}
	\caption{Multiscale statistical validation for the synthetic fast solar wind. Top-left panel highlights the spectral break (vertical blue dashed line) separating the $f^{-1}$ injection range from the $f^{-5/3}$ inertial range, indicated by red and green dashed lines, respectively. Top-right panel displays the characteristically low magnetic compressibility, with an average of $0.237$ over the entire frequency range (shown as a horizontal red dashed line), validating the highly Alfv\'enic nature of the fast wind. Bottom panel shows the scale-dependent flatness $\mathcal{F}(\tau)$, capturing the expected non-Gaussian small-scale intermittency (the horizontal blue dashed line marks the $\mathcal{F}=3$ Gaussian level).}
	\label{fig:fast_wind_spectra}
\end{figure}

The multiscale statistical analysis in Figure \ref{fig:fast_wind_spectra} confirms the profound physical differences, with respect to the slow regime, captured by the mathematical framework.

Unlike the slow wind, the fast wind PSD (top-left panel) exhibits a clear spectral break occurring at approximately $\sim1$ hour. This break mathematically separates the low-frequency energy injection range, characterized by a $f^{-1}$ spectral slope, from the Kolmogorov-like inertial range, scaling as $f^{-5/3}$. The magnetic compressibility $C_{B}(f)$ (top-right panel) is remarkably low across the inertial range, providing rigorous numerical evidence that the synthetic fast wind maintains the highly Alfv\'enic, nearly incompressible fluctuations observed in situ. Finally, the flatness $\mathcal{F}(\tau)$ (bottom panel) correctly increases towards the smallest scales, proving that the multiscale intermittency generator is robust and active independent of the background macroscopic state.

\subsection{Extreme transients: Coronal Mass Ejections}
\label{subsec:experiment_cme}
The final and most challenging numerical experiment evaluates the framework's capability to model extreme nonlinear transients, specifically the propagation of a CME embedded within a highly Alfv\'enic fast solar wind stream. The exhaustive set of macroscopic, topological, and stochastic parameters used to strictly initialize both the ambient medium and the transient structure at $1$ au is detailed in Table \ref{tab:cme_params}.

\begin{table}[ht]
	\centering
	\begin{tabular}{lc}
		\hline
		\hline
		\multicolumn{2}{c}{\textbf{Fast solar wind parameters}}\\
		\hline
		Bulk speed [km s$^{-1}$] & $800$\\
		Proton density [cm$^{-3}$] & $3$\\
		Proton temperature [K] & $2.5\times10^{5}$\\
		Mean magnetic field [nT] & $7.5$\\[2ex]
		\multicolumn{2}{c}{\textbf{Turbulent properties}}\\
		\hline
		Alfv\'enicity & $0.9$\\
		Magnetic compressibility & $0.1$\\[2ex]
		\multicolumn{2}{c}{\textbf{CME macroscopic properties}}\\
		\hline
		Max magnetic amplitude [nT] & $5$\\
		Max temporal duration [h] & $15$\\
		Radial expansion speed [km s$^{-1}$] & $40$\\
		Density drop (fraction) & $0.5$\\
		Temperature drop (fraction) & $0.7$\\[2ex]
		\multicolumn{2}{c}{\textbf{Internal flux rope \& shock parameters}}\\
		\hline
		Tilt angle $\theta$ [$^{\circ}$] & $90$\\
		Azimuthal angle $\phi$ [$^{\circ}$] & $90$\\
		Magnetic asymmetry factor & $0$\\
		Shock velocity jump scale [km/s] & $80$\\
		Shock/CME temporal gap [h] & $3$\\
		Shock driving efficiency & $0.3$\\[2ex]
		\multicolumn{2}{c}{\textbf{Sheath region turbulence parameters}}\\
		\hline
		Density fluctuation scale & $1.5$\\
		Bulk heating factor & $1.1$\\
		Thermal fluctuation amplitude [K] & $2\times10^{4}$\\
		Parallel velocity turbulence [km s$^{-1}$] & $40$\\
		Transverse velocity turbulence [km s$^{-1}$] & $20$\\[2ex]
		\multicolumn{2}{c}{\textbf{CME impact geometry parameters}}\\
		\hline
		 Longitude (Sun-Earth direction) [$^{\circ}$] & $5$\\
		 Latitude (ecliptic elevation) [$^{\circ}$] & $3$\\
		 Angular width [$^{\circ}$] & $30$\\
		\hline
		\hline
	\end{tabular}
	\caption{Comprehensive initialization setup for the CME numerical experiment, specifying the steady-state fast wind conditions, macroscopic CME geometry, flux rope properties, and the highly turbulent sheath amplification factors.}
	\label{tab:cme_params}
\end{table}

While the underlying mathematical model computes the exact $3$D topology across all arbitrary distributed coordinates continuously, visualizing the resulting multi-point time series for such a complex internal structure would result in severe graphical clutter. Therefore, to clearly appreciate the exact topological rotation of the magnetic flux rope and the thermodynamic stratification of the surrounding plasma, the synthetic time series is extracted on the ecliptic plane at a single specific observational node.

The dynamic passage of the CME is characterized by two distinct macroscopic sub-structures: the turbulent sheath region, bounded by a leading fast-forward shock, and the inner MC, modeled via the exact analytical force-free flux rope solutions defined in \S{} \ref{subsec:cme_modeling}.

Figure \ref{fig:cme_magnetic} displays the four-panel temporal evolution of the magnetic field vector.

\begin{figure}[ht]
	\centering
	\includegraphics[width=0.8\textwidth]{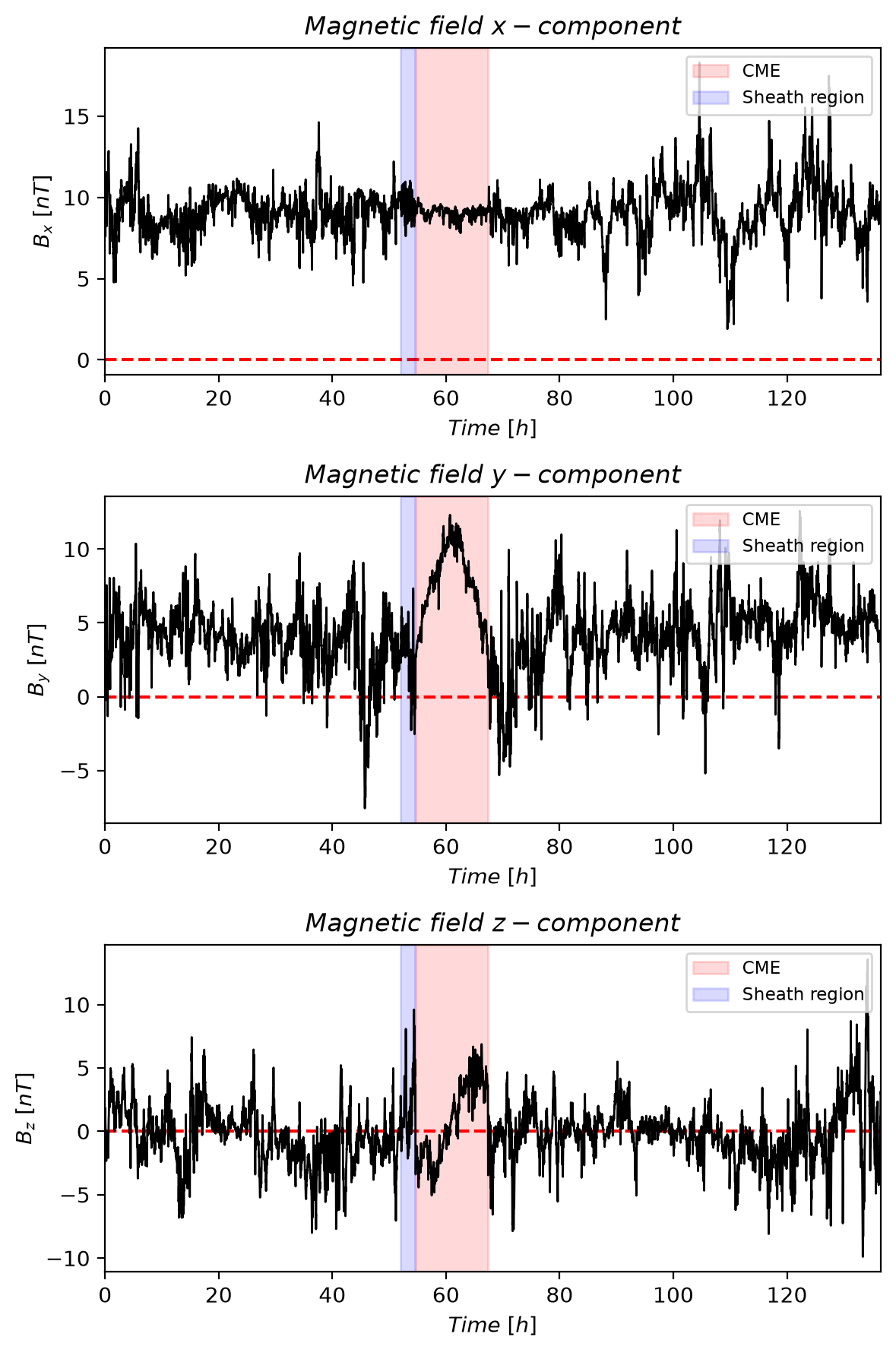}
	\caption{Synthetic magnetic field time series of a CME embedded in a fast solar wind stream, observed at a single ecliptic node. Panels from top to bottom display the magnetic field components $B_{x}$, $B_{y}$, $B_{z}$, and the total magnitude $\vert{}B\vert{}$. The shaded areas delineate the highly fluctuating sheath region (violet) and the coherent MC (pink). The topological rotation of the force-free flux rope is clearly visible in the vector components perpendicular to the sampling (radial) direction within the MC, with the horizontal red dashed line indicating the $0$ nT level to highlight the bipolar zero-crossing.}
	\label{fig:cme_magnetic}
\end{figure}

\begin{figure}[ht]
	\ContinuedFloat
	\centering
	\includegraphics[width=0.8\textwidth]{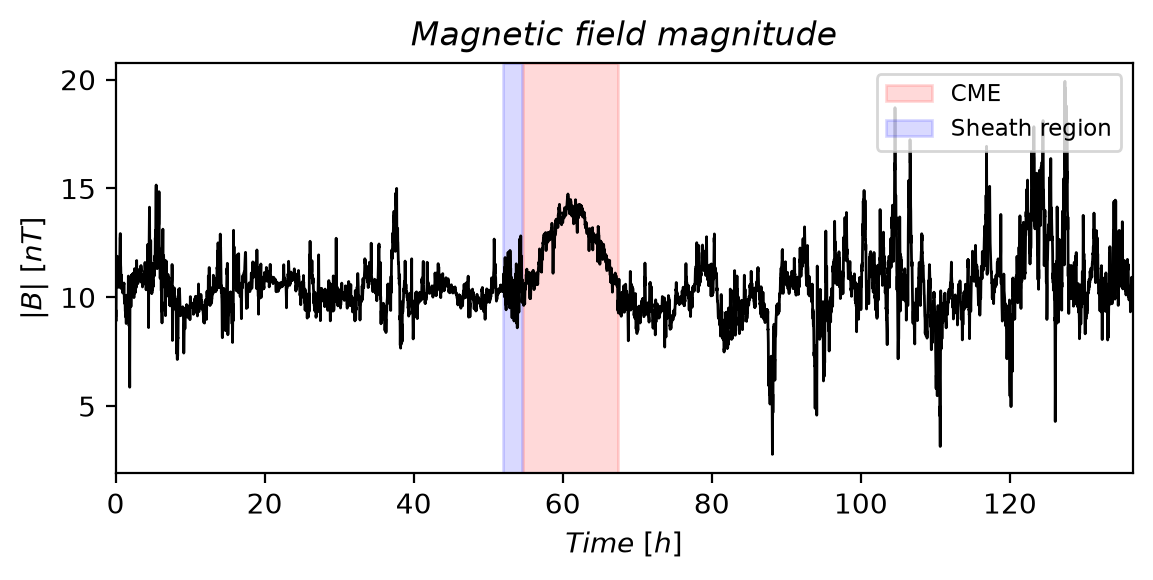}
\caption{\textit{(continued)}}
	\label{fig:cme_magnetic}
\end{figure}

The arrival of the transient is unequivocally marked by an overall enhancement of the magnetic field magnitude $\vert{}B\vert{}$ (bottom panel), signifying the passage of the collisionless forward shock. Although the macroscopic shock is mathematically formalized as a discrete Rankine-Hugoniot discontinuity, the synthetic temporal profile realistically avoids an idealized step function. Instead, the underlying geometric jump is immediately masked by the massive amplification of the superimposed stochastic cascade. Consequently, the sheath region exhibits highly amplified, strongly compressive turbulent fluctuations, accurately reflecting the stochastic processing of the ambient plasma dynamically compressed by the expanding ejecta. In stark contrast, as the virtual spacecraft enters the shaded MC region, the high-frequency turbulent noise is intrinsically suppressed. Specifically, the mathematical framework dynamically couples the prescribed variance of the stochastic generator to the macroscopic magnetic topology via the quenching operator $\eta_{quench}$ (see \S{} \ref{subsec:cme_modeling}), explicitly shutting down the turbulent multiplicative cascade where the coherent internal magnetic pressure strictly dominates the ambient low-$\beta$ ejecta. Within this region, the vector components perpendicular to the CME propagation direction ($B_{y}$ and $B_{z}$, middle two panels) exhibit the distinct smooth, large-scale coherent rotation mathematically dictated by the asymmetric, force-free Lundquist-derived topology.

Figure \ref{fig:cme_plasma} complements the MHD picture by presenting the six-panel temporal evolution of the plasma kinematics and thermodynamics.

\begin{figure}[ht]
	\centering
	\includegraphics[width=0.8\textwidth]{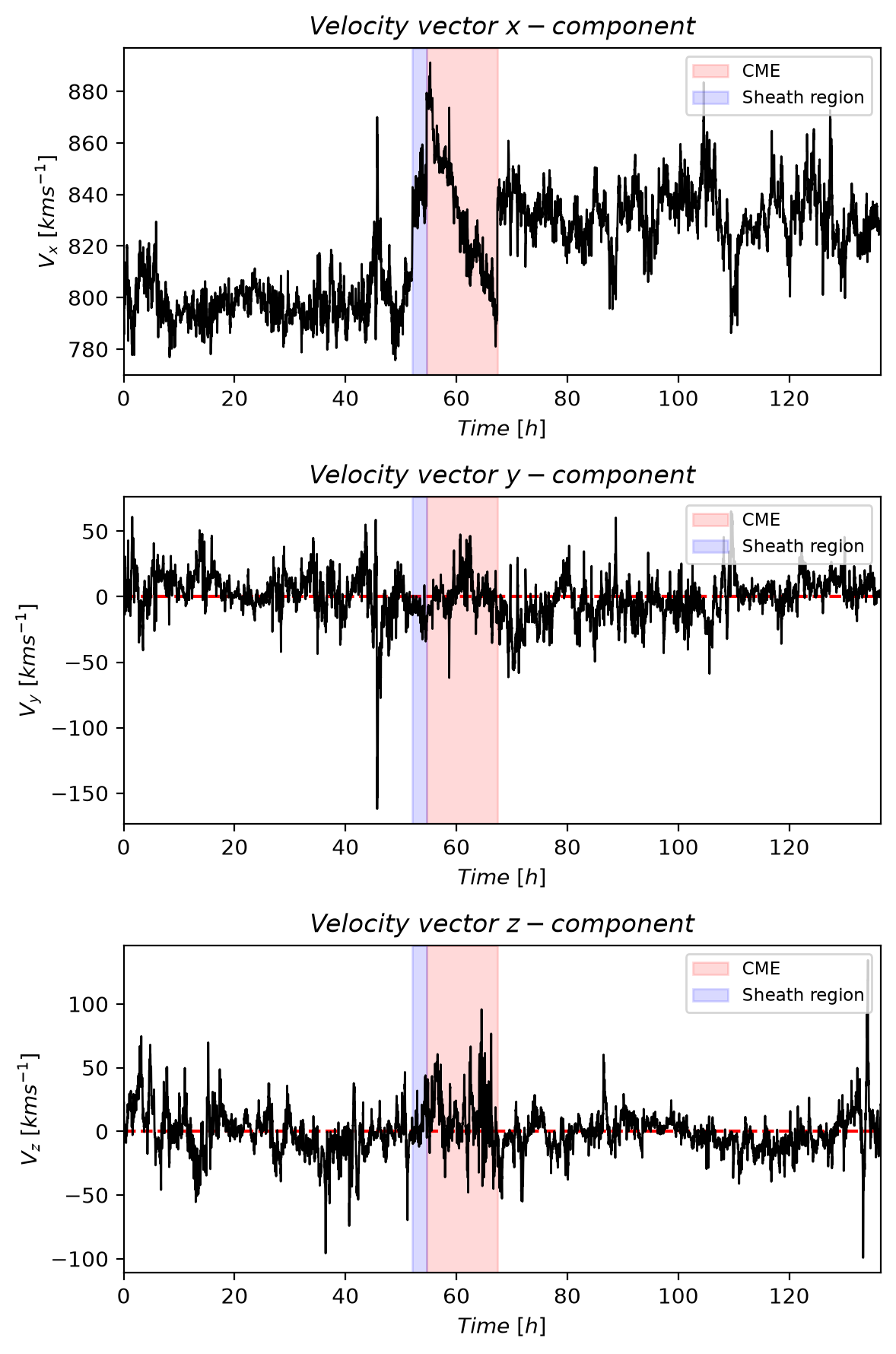}
	\caption{Synthetic plasma time series for the same CME event shown in Figure \ref{fig:cme_magnetic}. The six panels display the velocity components $V_{x}$, $V_{y}$, $V_{z}$, the bulk velocity magnitude $\vert{}V\vert{}$, the proton density, and temperature. The horizontal red dashed line in the $V_{y}$ and $V_{z}$ panels indicates the $0$ km s$^{-1}$ level, guiding the eye for the transverse velocity components that typically fluctuate around zero in the solar wind. The violet and pink shaded regions indicate the compressed, heated sheath and the expanding, cooling MC, respectively.}
	\label{fig:cme_plasma}
\end{figure}

\begin{figure}[ht]
	\ContinuedFloat
	\centering
	\includegraphics[width=0.8\textwidth]{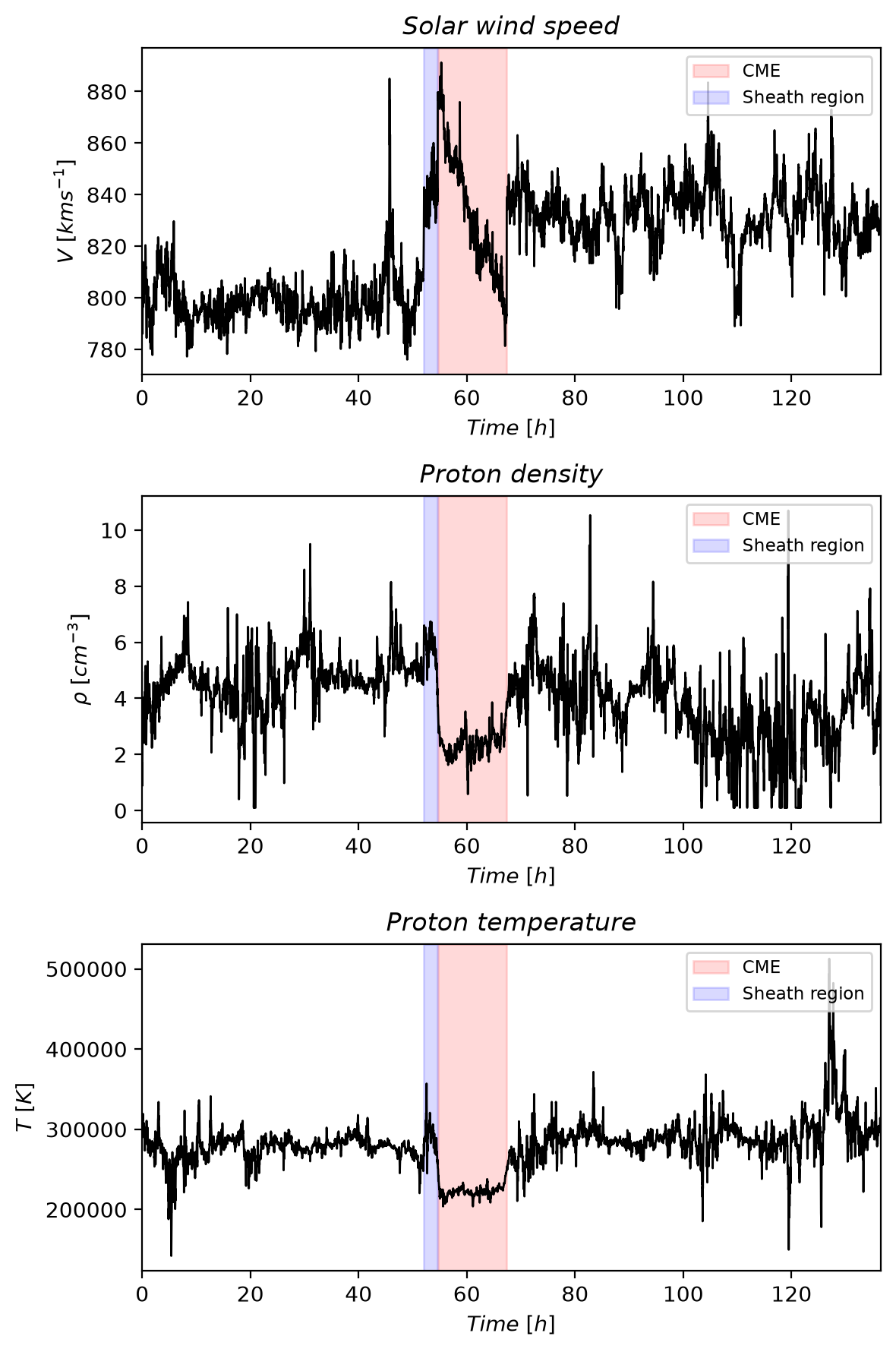}
\caption{\textit{(continued)}}
	\label{fig:cme_plasma}
\end{figure}

The kinematic and thermodynamic coupling within the $2.5$D framework proves exceptionally robust. The forward shock induces an instantaneous, discontinuous jump in the bulk velocity $\vert{}V\vert{}$, density, and temperature. The sheath region acts as a highly dense and anomalously hot buffer zone. Upon crossing into the MC, the framework naturally enforces the expected thermodynamic signatures of a rapidly expanding coherent structure: the proton density and temperature drop significantly below the ambient fast wind values, preserving the exact local pressure balances. Furthermore, the total velocity magnitude $\vert{}V\vert{}$ within the MC exhibits a strictly linear decline, which is the exact mathematical signature of a radially expanding kinematic structure conserving its self-similar geometrical properties as it propagates outwardly.

In summary, these numerical experiments demonstrate that the proposed mathematical framework successfully bridges the fundamental scale-separation gap. It harmoniously integrates the exact analytical formulations of macroscopic heliospheric structures (CIRs, HCSs, and CMEs) with the rigorous statistical constraints of fully developed stochastic turbulence, providing a computationally efficient and physically exact baseline for multi-point heliospheric modeling.

\section{Concluding remarks}
\label{sec:conclusions}
The accurate modeling of the inner heliosphere is historically hindered by a severe scale-separation problem, wherein macroscopic coherent structures propagate or are advected through a background medium intrinsically dominated by fully developed, highly nonlinear stochastic turbulence. In this paper, a novel mathematical and numerical framework designed to harmoniously bridge this fundamental gap has been introduced. By circumventing the computationally prohibitive constraints of brute-force global $3$D numerical solvers in favor of exact analytical formulations coupled with a discrete multiplicative cascade (p-model), a highly realistic, synthetic continuous heliospheric environment has been successfully generated.


The comprehensive numerical experiments demonstrate the profound physical fidelity of this unified approach. The framework dynamically evolves massive macroscopic topologies, such as kinematic steepening in Corotating Interaction Regions, continuous sector reversals across the Heliospheric Current Sheet, and extreme, force-free Coronal Mass Ejections, while rigorously preserving the small-scale theoretical constraints of the ambient plasma. The model has been shown to exert strict analytical control over the spectral energy distribution, correctly replicating the highly compressible $f^{-5/3}$ inertial range of the slow solar wind, the $f^{-1}$ injection break of the highly Alfv\'enic fast wind, and the non-Gaussian intermittency signatures dictated by the scale-dependent Flatness.

Beyond its theoretical elegance, the most significant implication of this multiscale model lies in its operational capability to support the architectural design and feasibility studies of future in-situ multi-spacecraft observatories. The transition from single-point L$1$ monitors to distributed observational networks, such as the SHIELD mission concept currently under investigation by the European Space Agency (ESA), is a mandatory evolutionary step to finally break the spatial degeneracy inherent in $1$D Eulerian temporal cuts. By generating exact spatio-temporal arrays across arbitrary distributed coordinates, the present framework provides a computationally highly efficient, stochastically perturbed synthetic testbed.

This environment provides mission planners with the unprecedented ability to optimize the geometric configuration of a spacecraft swarm \textit{a priori}. It allows the rigorous assessment of how varying radial and azimuthal baselines affect the geometric triangulation of shock fronts, the extraction of propagation vectors, and the $3$D reconstruction of magnetic flux ropes. Ultimately, validating multi-point inversion algorithms and data assimilation techniques requires a synthetic environment that exhibits both the macroscopic curvature of the transient and the statistically decorrelated turbulent noise unique to each observational node. By satisfying both criteria simultaneously, this mathematical framework establishes a robust analytical baseline for computational space physics, directly paving the way for the next generation of distributed heliospheric exploration.

\bibliographystyle{elsarticle-harv}
\bibliography{reference_list}

\end{document}